\documentclass[letterpaper,twocolumn,prl,aps,superscriptaddress,floatfix]{revtex4}

\usepackage[latin9]{inputenc}
\usepackage{amsmath}
\usepackage{amssymb}
\usepackage{graphicx}
\usepackage{esint}
\usepackage{times}
\usepackage[markup=underlined]{changes}

\usepackage{comment}
\usepackage{upgreek}

\usepackage[unicode=true,
bookmarks=true,bookmarksnumbered=false,bookmarksopen=false,
breaklinks=false,pdfborder={0 0 1},backref=false,colorlinks=true]
{hyperref}
\hypersetup{
    linkcolor=magenta, urlcolor=blue, citecolor=blue, pdfstartview={FitH}, hyperfootnotes=false, unicode=true}

\makeatletter

\pdfpageheight\paperheight
\pdfpagewidth\paperwidth

\@ifundefined{textcolor}{}
{%
    \definecolor{BLACK}{gray}{0}
    \definecolor{WHITE}{gray}{1}
    \definecolor{RED}{rgb}{1,0,0}
    \definecolor{GREEN}{rgb}{0,1,0}
    \definecolor{BLUE}{rgb}{0,0,1}
    \definecolor{CYAN}{cmyk}{1,0,0,0}
    \definecolor{MAGENTA}{cmyk}{0,1,0,0}
    \definecolor{YELLOW}{cmyk}{0,0,1,0}
}

\usepackage{xcolor}
\usepackage{soul}

\newcommand{\ket}[1]{\ensuremath{\left|#1\right\rangle}}
\definecolor{blue}{rgb}{0,0,1}
\definecolor{red}{rgb}{1,0,0}
\definecolor{green}{rgb}{0,1,0}

\makeatother

\begin{document}
\title{Quantum-enhanced dark matter search using cat states}

\author{Pan Zheng}
\thanks{These authors contributed equally to this work.}
\affiliation{International Quantum Academy, Shenzhen 518048, China}

\author{Yanyan Cai}
\thanks{These authors contributed equally to this work.}
\affiliation{International Quantum Academy, Shenzhen 518048, China}
\affiliation{Southern University of Science and Technology, Shenzhen 518055, China}

\author{Bin Xu}
\thanks{These authors contributed equally to this work.}
\affiliation{School of Physics and State Key Laboratory of Nuclear Physics and Technology, Peking University, Beijing 100871, China}

\author{Shengcheng Wen}
\author{Libo Zhang}
\affiliation{International Quantum Academy, Shenzhen 518048, China}
\affiliation{Southern University of Science and Technology, Shenzhen 518055, China}

\author{Zhongchu Ni}
\affiliation{International Quantum Academy, Shenzhen 518048, China}

\author{Jiasheng Mai}
\affiliation{International Quantum Academy, Shenzhen 518048, China}
\affiliation{Southern University of Science and Technology, Shenzhen 518055, China}

\author{Yanjie Zeng}
\affiliation{
Institute of Theoretical Physics, Chinese Academy of Sciences, Beijing 100190, China}
\affiliation{
School of Physical Sciences, University of Chinese Academy of Sciences, Beijing 100049, China}
\author{Lin Lin}
\affiliation{School of Physics and State Key Laboratory of Nuclear Physics and Technology, Peking University, Beijing 100871, China}
\affiliation{Institute of Heavy Ion Physics, Peking
University, Beijing 100871, China}

\author{Ling Hu}
\affiliation{International Quantum Academy, Shenzhen 518048, China}
\affiliation{Shenzhen Branch, Hefei National Laboratory, Shenzhen 518048, China}
\author{Xiaowei Deng}
\affiliation{International Quantum Academy, Shenzhen 518048, China}

\author{Song Liu}
\email{lius3@sustech.edu.cn}
\affiliation{International Quantum Academy, Shenzhen 518048, China}
\affiliation{Shenzhen Branch, Hefei National Laboratory, Shenzhen 518048, China}
\author{Jing Shu}
\email{jshu@pku.edu.cn}
\affiliation{School of Physics and State Key Laboratory of Nuclear Physics and Technology, Peking University, Beijing 100871, China}
\affiliation{Center for High Energy Physics, Peking University, Beijing 100871, China}
\affiliation{Beijing Laser Acceleration Innovation Center, Huairou, Beijing, 101400, China}
\author{Yuan Xu}
\email{xuyuan@iqasz.cn}
\affiliation{International Quantum Academy, Shenzhen 518048, China}
\affiliation{Shenzhen Branch, Hefei National Laboratory, Shenzhen 518048, China}
\author{Dapeng Yu}
\email{yudapeng@iqasz.cn}
\affiliation{International Quantum Academy, Shenzhen 518048, China}
\affiliation{Shenzhen Branch, Hefei National Laboratory, Shenzhen 518048, China}

\begin{abstract}
Quantum metrology has recently emerged as a powerful approach for dark matter (DM) searches, particularly using nonclassical bosonic states in microwave cavities that are sensitive to weak signals. Nonclassical cat states---macroscopic superpositions of coherent states featuring sub-Planck interference structures---offer promising advantages for high-precision measurements. However, their practical utility in DM search remains unexplored. Here, we report the first experimental application of four-component cat states within a high-quality superconducting microwave cavity to search for dark photons, a potential DM candidate. We demonstrate an 8.1-fold enhancement in the signal photon rate and constrain the dark photon kinetic mixing angle to an unprecedented $\epsilon < 7.32 \times 10^{-16}$ near 6.44~GHz (26.6~$\upmu$eV). By employing a parametric sideband drive to actively tune the cavity frequency, we achieve dark photon searches and background subtraction across multiple frequency bins, yielding a sensitivity at the $10^{-16}$ level within a 100~kHz bandwidth. Our cat-assisted DM (CaD) search and frequency-scanning techniques demonstrate substantial improvements over previous results, promising potential implications in quantum-enhanced searches for new physics.
\end{abstract}

\maketitle
%\section{Introduction}
Exploring the nature of dark matter (DM) remains one of the most pressing challenges in modern fundamental physics. Although numerous observations in astronomy~\cite{Sofue2001, Massey2010} and cosmology~\cite{Markevitch2004} confirm the existence of DM, its fundamental properties remain mysterious. Among various DM candidates, ultra-light bosons, particularly axions~\cite{Preskill1982, Abbott1982, Dine1982} and dark photons (DPs)~\cite{Holdom1986, Nelson2011}, stand out as they naturally arise in many extensions of fundamental physics beyond the Standard Model~\cite{Peccei:1977hh, Svrcek2006, Abel2008, Arvanitaki2009, Goodsell2009}.

Haloscope experiments are underway to detect these bosonic DM candidates by exploiting their weak couplings to electromagnetic fields. In the microwave regime, cavities are employed to detect weak signals arising from axions via their conversion in a strong magnetic field~\cite{Sikivie1983, McAllister:2017lkb, ADMX:2019uok, CAPP:2020utb}, or from DPs via direct coupling to photons in superconducting cavities~\cite{Wagner2010, Ghosh2021, Cervantes:2022gtv,SHANHE:2023kxz,Kang:2024slu}. The critical challenge lies in distinguishing these extremely weak signals from noise, including both thermal noise and quantum fluctuations, necessitating unprecedented detection sensitivity.

Quantum metrology can significantly enhance the detection sensitivity beyond the standard quantum limit (SQL) by employing quantum squeezed states~\cite{backes2021}, mode entanglement and state swapping techniques~\cite{Chen:2021bgy, Jiang:2022vpm}, and quantum non-demolition (QND) measurements with vacuum~\cite{Dixit2021} and Fock states~\cite{Agrawal2024}. Besides these advanced states, cat states~\cite{Schleich1991, Gerry1997}, which are quantum superpositions of quasiclassical coherent states, stand out due to their remarkable non-Gaussian features and sub-Planck interference structures in phase space~\cite{Zurek2001}. These unique characteristics enable cat states to improve measurement precision below the SQL~\cite{Munro2002, Pan2025}, distinguishing them from conventional coherent states. Despite their metrological potential, the practical use of cat states in DM search remains unexplored.

In this study, we present the first experimental demonstration of DPDM search using four-component cat states in a high-quality superconducting microwave cavity. The cavity's initial cat states are prepared via a dispersively coupled superconducting qubit, which also enables single-photon counting and error suppression through repeated QND measurements of the photon-number parity. By increasing the average photon numbers of the cat states, we achieve a maximum enhancement factor of 8.1 in the signal photon rate compared to the vacuum state and constrain the DP kinetic mixing angle $\epsilon < 7.32 \times 10^{-16}$ at the 90\% confidence level (C.L.) near 6.44~GHz (26.6~$\upmu$eV). Furthermore, through parametric sideband driving to actively tune the cavity frequency and subtracting the background noise across various frequency bins, we ultimately constrain the DP kinetic mixing angle on the order of $10^{-16}$ within a bandwidth of about 100~kHz, demonstrating the viability of quantum techniques for next-generation DM searches.

\begin{figure*}
	\includegraphics{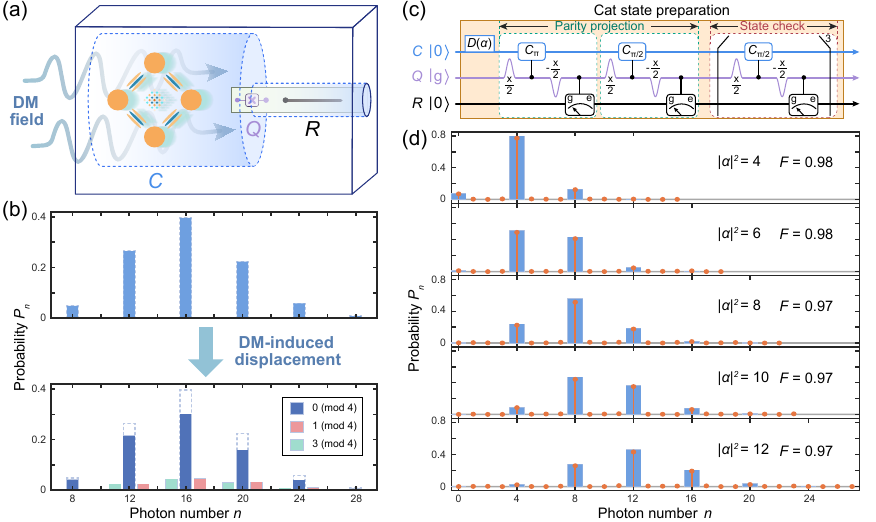}
	\caption{Illustration of the DP search scheme with cat states and state preparation results.
    (a) Schematic of the experimental device for DP search using compass states. 
    (b) The photon number distributions of a compass state before and after the DM-induced displacement operation.
    (c) Experimental sequence for preparing the compass states with two successive sinusoidal photon number filters and three parity checks to ensure successful state preparation. Here, $C_\theta=e^{i\theta a^\dagger a}$ represents the cavity phase shift of $\theta$ conditional on the qubit state.
    (d) Measured (dots) and ideal (bars) photon number populations of the generated compass states for different average photon numbers $|\alpha|^2$.  
    }
    \label{fig1}
\end{figure*}

%\section{Dark photon detection scheme using cat states}

The wave-like bosonic DPDM manifests itself as a coherent oscillating microwave field that weakly couples to Standard Model photons in a cavity via the kinetic mixing parameter $\epsilon$. Under the resonance condition, the DPDM field acts as a displacement operator $ D(\beta) = \exp(\beta a^\dagger - \beta^* a)$ within the DM coherence time $\tau_{\mathrm{DM}}$, where $\beta$ is proportional to $\epsilon$ and the signal integration time $\tau$, and $a^\dagger$, $a$ are the cavity ladder operators. The weak coupling leads to a small displacement ($|\beta| \ll 1$), which converts a vacuum state into a superposition $|\psi\rangle \propto |0\rangle + \beta |1\rangle + \mathcal{O}(\beta^2)$. By measuring the single-photon excitation probability $P_1 = |\langle 1 | D(\beta) | 0\rangle |^2\approx |\beta|^2$, we can extract $\beta$ and thus constrain $\epsilon$.

To enhance the DM-induced signal, the cavity can be prepared in nonclassical states, such as cat states~\cite{Schleich1991, Gerry1997}. 
While conventional two-component cat states---superpositions of phase-opposed coherent states $|\phi_\pm\rangle \propto |\alpha\rangle \pm |-\alpha\rangle$---have shown advantages in quantum metrology and information processing~\cite{Munro2002, Vlastakis2013, Ofek2016, Grimm2020, Reglade2024, Pan2025}, their susceptibility to single-photon loss introduces a critical limitation. 
Specifically, both the DM-induced displacement and single-photon loss lead to parity flipping between even ($|\phi_+\rangle$) and odd ($|\phi_-\rangle$) cat states, making it difficult to distinguish the DM-induced signals from relaxation errors in background noise. 

To address this challenge, we propose to use a four-component cat state, also known as a compass state~\cite{Zurek2001}, defined as 
\begin{equation}
|\phi_0(\alpha)\rangle \propto |\alpha\rangle +  |-\alpha\rangle + |i\alpha\rangle +|-i\alpha\rangle    
\end{equation}
for DM search. This state exhibits fourfold rotational symmetry and ultrafine interference structures in phase space, making it extremely sensitive to small displacements induced by the DM field, as illustrated in Fig.~\ref{fig1}(a). The photon number occupations of this state follow a Poisson distribution restricted to $n = 0\,( \mathrm{mod}\,4)$, referred to as ``doubly even" parity. Figure~\ref{fig1}(b) depicts the photon number distributions of the compass state $|\phi_0(\alpha)\rangle$, which evolves into a superposition state $|\psi\rangle \propto |\phi_0(\alpha)\rangle + \beta\alpha^*|\phi_1(\alpha)\rangle - \beta^*\alpha|\phi_3(\alpha)\rangle +\mathcal{O}(\beta^2)$ under a weak DM-induced displacement ($|\beta| \ll 1$). Here, $|\phi_{1/3}(\alpha)\rangle \propto |\alpha\rangle - |-\alpha\rangle \mp i |i\alpha\rangle  \pm i|-i\alpha\rangle$ represents the left- and right-odd parity cat states with photon numbers restricted to $n=1\,( \mathrm{mod}\,4)$ and $n=3\,( \mathrm{mod}\,4)$ subspaces, respectively. 
We project the evolved cat state onto the ``left-odd" parity state $|\phi_1(\alpha)\rangle$, with the projection probability given by 
\begin{equation}
\label{eq:compassprobility}
P_\alpha(\beta) = \left|\langle \phi_1(\alpha)|D(\beta)|\phi_0(\alpha)\rangle\right|^2 
\approx |\beta|^2 |\alpha|^2,
\end{equation}
thereby demonstrating a signal enhancement by a factor of $|\alpha|^2$ compared to the vacuum-state-based scheme. This approach provides an inherent error discrimination since single-photon loss converts $|\phi_0(\alpha)\rangle$ to $|\phi_3(\alpha)\rangle$ rather than $|\phi_1(\alpha)\rangle$. This property enables robust separation of DM-induced signals from single-photon-loss errors in the background noise through generalized modulo-4 photon-number-parity measurements.

\begin{figure*}
    \includegraphics{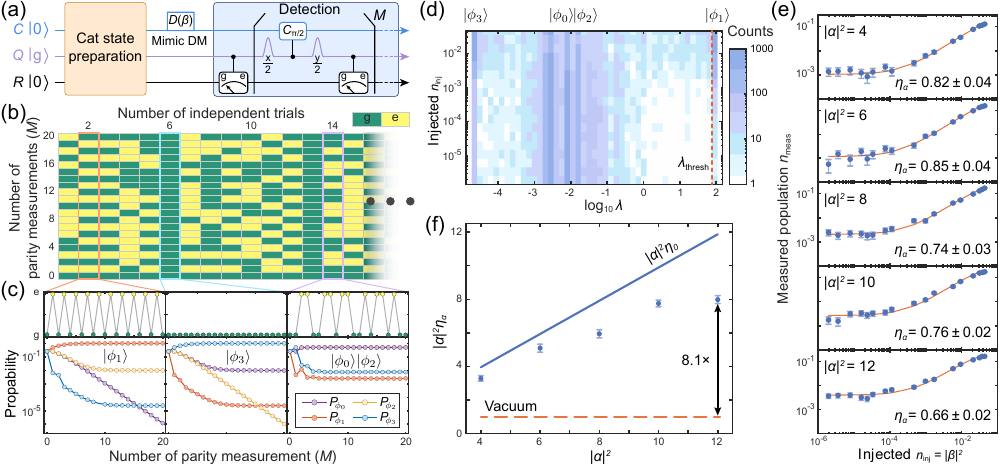}
    \caption{Characterization of the DM detector.
(a) Experimental sequence for characterizing the DM detector. A displacement operator is applied to mimic the effect of the DM wave, and $M$ repetitions of QND parity measurements are performed to extract the cavity state probabilities within different subspaces.
(b) Measured single-shot readout signals of the $M=20$ repetitions of the parity measurements for different independent trials. 
(c) Three vertical cuts of readout signals in (b) (top row) and the corresponding reconstructed state probabilities $P_{\phi_i}$ (bottom row) within $|\phi_i\rangle$ subspace via the hidden Markov model analysis.
(d) Histograms of likelihood ratios $\lambda = {P_{\phi_1}}/{(P_{\phi_0}+P_{\phi_2}+P_{\phi_3})}$ of all events for various injected mean photon numbers $n_\mathrm{inj}=|\beta|^2$. The dashed red line indicates the chosen threshold $\lambda_{\rm{thresh}}=84$.  
(e) Statistical probabilities (dots) of $\lambda>\lambda_\mathrm{thresh}$ events as a function of $n_\mathrm{inj}$ for compass states with various $|\alpha|^2$, which are extracted from (d). Detector efficiency ($\eta_\alpha$) is determined from the linear fits (solid red lines).
(f) The total detection efficiency $|\alpha|^2\eta_\alpha$ [extracted from the fits in (e)] for $|\phi_0(\alpha)\rangle$ as a function of $|\alpha|^2$, demonstrating a maximum enhancement factor of 8.1 compared to the vacuum state (dashed red line). 
    }
    \label{fig2}
\end{figure*}

It is noted that although the $|\alpha|^2$-enhancement of the compass state comes with a reduction factor of $1/|\alpha|^2$ in its coherence lifetime, the signal rate remains unaffected. This is because the coherence time is primarily limited by the DM bandwidth rather than by the compass-state lifetimes, given the cavity's long coherence time. This mechanism is similar to that of the Fock-state-based DM search scheme~\cite{Agrawal2024}.

%\section{Cat states preparation and detector characterization}

In our experiment, the probe cavity for searching for the DP signals is constructed by using the $\rm TM_{010}$ mode of a cylindrical niobium cavity, which has a resonance frequency $\omega_c = 2\pi \times 6.442 ~\text{GHz}$ and a quality factor $Q \sim 1.86 \times 10^8$ ($T_1^c=4.6$~ms). The cavity is dispersively coupled to a superconducting transmon qubit ($\omega_q = 2\pi \times 5.205 ~\text{GHz}$ and $T_1^q=175~\upmu$s) with a Hamiltonian $H = \omega_c a^\dagger a + \omega_q|e\rangle\langle e| - \chi_{qc} a^\dagger a |e\rangle\langle e|$, enabling control and measurement of the cavity states. Here, $|e\rangle$ ($|g\rangle$) denotes the first excited (ground) state of the qubit, and $ \chi_{qc} = 2\pi \times 0.6~\text{MHz}$ represents the dispersive coupling strength between the qubit and cavity. Additionally, the qubit is connected to a stripline readout resonator ($\omega_r = 2\pi \times 7.950~\text{GHz}$) for measuring qubit states. A schematic of the experimental device is depicted in Fig.~\ref{fig1}(a).

\begin{figure*}
    \includegraphics{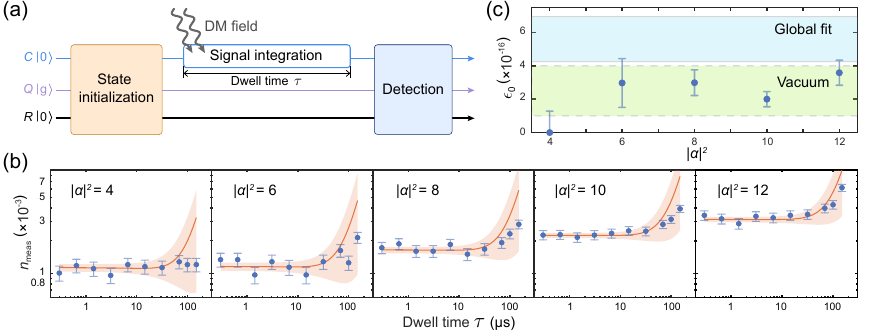}
    \caption{DP search using compass states.
	 (a) Experimental sequence for conducting DP searches as varying the dwell time $\tau$.
    (b) Measured probabilities of positive events as a function of $\tau$ for compass states with different amplitudes $|\alpha|^2$. Solid lines and red regions represent the global fitting to Eq.~\eqref{eq:fit_a0}, and the 90\% C.L., respectively.
    (c) Extracted kinetic mixing angle $\epsilon_0 \pm \sigma_{\epsilon_0}$ versus $|\alpha|^2$ from individual fits in (b), which is compared to that from the global fitting with all the compass states (blue region) and that from the vacuum-state-based scheme (green region).
}
\label{fig3}
\end{figure*}

The dispersive Hamiltonian induces a photon-number-dependent qubit frequency shift~\cite{Schuster2007}, enabling QND measurements of cavity photons~\cite{Sun2014} and projections of cavity states into generalized photon-number parities~\cite{Vlastakis2013, Deng2024}. Using this dispersive interaction, we demonstrate the preparation of the initial compass state $|\phi_0(\alpha)\rangle$ in the cavity for DM searches. This is achieved through the sequence illustrated in Fig.~\ref{fig1}(c). First, the cavity is initialized in a coherent state $|\alpha\rangle$. Subsequently, two successive sinusoidal photon number filter operations~\cite{Deng2024} are applied to project the coherent state into a doubly even parity subspace after postselecting qubit ground states in both filters. Finally, three parity checks are performed to ensure minimal photon number occupations outside the $|\phi_0(\alpha)\rangle$ subspace. The generated compass states are characterized via qubit spectroscopy experiments to extract their photon-number populations, as shown in Fig.~\ref{fig1}(d). The performance of the compass state preparation is quantified by calculating the statistical overlapping~\cite{fuchs1996} fidelity $F=\sum_n{\sqrt{P_n^\mathrm{meas}P_n^\mathrm{ideal}}}$ between the measured and ideal photon number populations of the compass states. The results indicate that the fidelities of the compass states exceed 95\% with average photon numbers up to $|\alpha|^2=12$.

After preparing the initial cat states within the cavity, we then characterize the DP detector using the experimental sequence shown in Fig.~\ref{fig2}(a). In this sequence, we apply a displacement operation $D(\beta)$ to the initial compass state $|\phi_0(\alpha)\rangle$ to mimic the effect of the DM wave. We then perform 20 repetitions of QND parity measurements, similar to those used for initial-state preparation, to determine the cavity-state probabilities within different photon-number-parity subspaces. 
The measured single-shot results of the qubit states are presented in Fig.~\ref{fig2}(b), with three vertical cuts highlighted in Fig.~\ref{fig2}(c). These results are used to determine the probabilities $P_{\phi_3}$, $P_{\phi_1}$, and $P_{\phi_0+\phi_2}$ for cavity states within $|\phi_3(\alpha)\rangle$, $|\phi_1(\alpha)\rangle$, and \{$|\phi_0(\alpha)\rangle$, $|\phi_2(\alpha)\rangle$\} subspaces using a hidden Markov model analysis~\cite{Dixit2021, Agrawal2024,supplement}. Here, $|\phi_2(\alpha)\rangle \propto |\alpha\rangle + |-\alpha\rangle - |i\alpha\rangle - |-i\alpha\rangle$ represents the cat state with photon numbers restricted in $n=2\,( \mathrm{mod}\,4)$ subspace. We then calculate the likelihood ratio, $\lambda = {P_{\phi_1}}/{(P_{\phi_0+\phi_2}+P_{\phi_3})}$, to assess the transition probability from $|\phi_0(\alpha)\rangle$ to $|\phi_1(\alpha)\rangle$. A histogram of $\lambda$ for various injected average photon numbers $n_\mathrm{inj}=|\beta|^2$ is shown in Fig.~\ref{fig2}(d).

Here we set a likelihood threshold of $\lambda_{\mathrm{thresh}} = 84$ to discriminate positive and negative events of measuring the $|\phi_1(\alpha)\rangle$ state, with $\lambda > \lambda_{\mathrm{thresh}}$ ($\lambda \leq \lambda_{\mathrm{thresh}}$) indicating positive (negative) events. This threshold is chosen practically based on an optimal trade-off between suppressing false-positive events and maintaining high detection efficiency~\cite{supplement}. We then compute the statistical probability $n_\mathrm{meas}$ of these positive events, which is plotted as a function of $n_\mathrm{inj}=|\beta|^2$ for compass states $|\phi_0(\alpha)\rangle$ with different $|\alpha|^2$ in Fig.~\ref{fig2}(e). These traces are fitted to a model
\begin{equation}\label{eq:p01}
    n_{\rm meas}=\eta_\alpha P_\alpha(\beta)+\delta_\alpha,
\end{equation}
where $P_\alpha(\beta)=|\alpha|^2n_\mathrm{inj}$ represents the projection probability in Eq.~\eqref{eq:compassprobility}, and $\eta_\alpha$ and $\delta_\alpha$ denote the detection efficiency and false positive probability, respectively. The resulting overall detection efficiency $|\alpha|^2 \eta_\alpha $ is plotted as a function of $|\alpha|^2$ in Fig.~\ref{fig2}(f), demonstrating a maximum enhancement of $\eta_\alpha |\alpha|^2/\eta_0=8.1$ between the compass state $|\phi_0(\alpha=\sqrt{12})\rangle$ and the vacuum state. The decrease of the enhancement at large $|\alpha|$ is primarily due to additional state preparation and parity measurement errors, leading to increased false-positive DP signals from higher $|\phi_2(\alpha)\rangle$ populations~\cite{supplement}.

%\section{Dark photon search using cat states}

After characterizing the compass-state-based DM detector, we conduct a DP search experiment with the sequence shown in Fig.~\ref{fig3}(a). In this experiment, we replace the mimic DM drive with a signal-integration process between the state initialization and detection to coherently accumulate the signals induced by the real DM field. 
After preparing the cavity in $|\phi_0(\alpha)\rangle$ with different $\alpha$, we perform independent DP search experiments and calculate the probability of positive events $n_{\mathrm{meas}}$ as varying the signal integration time $\tau$. The measured probabilities $n_{\mathrm{meas}}$ are plotted as a function of $\tau$ in Fig.~\ref{fig3}(b) and fitted to the function:  
\begin{equation}
    n_{\mathrm{meas}} = a_0 \eta_{\alpha} |\alpha|^2 g(\tau) + b_\alpha\tau + c_{\alpha}.
    \label{eq:fit_a0}
\end{equation}
Here, the first term accounts for the contributions from the DM wave, with a time dependence factor of $g(\tau)$, which is determined by the DM lineshape and given by 
\begin{equation}
    g(t) \simeq \begin{cases}
        t^2  & \text{if } 0 < t < \tau_{\mathrm{DM}}, \\
        \tau_{\mathrm{DM}} t & \text{if } t > \tau_{\mathrm{DM}}.
    \end{cases}
\end{equation}
For $t > \tau_{\mathrm{DM}}$, $g(t)$ scales linearly with time, behaving as an incoherent source~\cite{supplement}. Therefore, we limit $\tau$ to be smaller than $\tau_{\mathrm{DM}}\approx 152 \, \upmu\mathrm{s}$ in this experiment.

\begin{figure*}
    \includegraphics{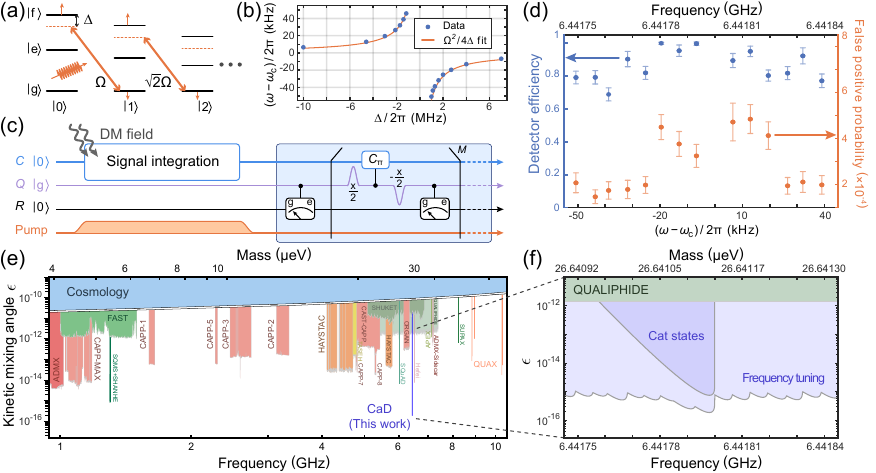}
    \caption{DP search while tuning the cavity frequency.
(a) Energy diagram of the cavity-qubit system, where the cavity frequency is tuned with a detuned parametric sideband drive.
(b) Measured cavity frequency relative to the original frequency $\omega_c$ without tuning as a function of the drive detuning $\Delta$.
(c) Experimental sequence for conducting the DPDM search at each cavity frequency with the cavity initialized in a vacuum state.
(d) Extracted detection efficiency and false positive probability of the cavity detector at each cavity frequency.
(e) Comparison of the exclusion of DPDM parameter space with previous results in the frequency range of 1-10~GHz. Our exclusion data is labeled as ``CaD" and highlighted in (f).
(f) DPDM exclusion results measured in this experiment for both the cat-state-based detection scheme and the frequency tuning method. 
    }
    \label{fig4}
\end{figure*}

For each compass state, we perform a Maximum Likelihood Estimation (MLE) on the measured probabilities to determine the fitting parameters ($a_0$, $b_{\alpha}$, and $c_{\alpha}$) and their uncertainties.
Using the fitted values of $a_0$, we can then calculate the DP kinetic mixing angle $\epsilon_0 = \sqrt{a_0/(\rho_{\mathrm{DM}} m_{\mathrm{DM}} V_{\rm eff})}$ with $\rho_{\mathrm{DM}} m_{\mathrm{DM}} V_{\rm eff} = 1.10 \times 10^{35} \, \mathrm{s}^{-2} $. Here $\rho_\mathrm{DM}$, $m_\mathrm{DM}$ represent the DM density and mass, and $V_\mathrm{eff}$ is the effective cavity volume. The uncertainty $\sigma_{\epsilon_0}$ is calculated using error propagation by taking into account the uncertainties in all parameters~\cite{supplement}. The resulting $\epsilon_0 \pm \sigma_{\epsilon_0}$, are plotted as a function of $|\alpha|^2$ in Fig.~\ref{fig3}(c) in comparison with that from the vacuum-state-based scheme~\cite{supplement}. A global MLE fit is also performed across all compass states using a common fitting parameter $a_0$. The global fit enforces a single, physically unique kinetic mixing angle shared across all compass states, yielding a more conservative uncertainty than individual fits. The resulting 90\% C.L. limit on $\epsilon$ from the global fit, is expressed as $\epsilon<\epsilon_0 + 1.28 \sigma_{\epsilon_0} = 7.32 \times 10^{-16}$ for DP on resonance with the cavity. The exclusion region near the resonance frequency is shaped by the lineshape of the DM field.

%\section{Dark photon search with frequency tuning}
Although the cat-state-based DP search scheme achieves an unprecedented sensitivity with on-resonance DP signals, the sensitivity to off-resonance signals drops quickly due to the finite bandwidth of the DM lineshape. To address this challenge, we propose dynamically in-situ tuning the cavity frequency and performing DP search experiments across various frequency bins to subtract the background noise. 

The cavity frequency tuning is achieved by using a detuned parametric sideband drive~\cite{Zeytinoglu2015, Rosenblum2018}, with the Hamiltonian described as
\begin{equation} 
H_d = \frac{\Omega}{2} a^\dagger |g\rangle \langle f| e^{i\Delta t} + \mathrm{h.c.}. 
\end{equation} 
Here, $\Omega$ denotes the drive strength, and $\Delta$ represents the frequency detuning between the drive frequency $\omega_d$ and the resonant frequency $\omega_{fg} - \omega_c$, where $\omega_{fg}$ is the frequency difference between the qubit ground state $|g\rangle$ and the second excited state $|f\rangle$. As illustrated in Fig.~\ref{fig4}(a), this drive couples the levels $|n,f\rangle$ to $|n+1,g\rangle$, with $n$ representing the photon number in the cavity.

For a sufficiently detuned sideband drive ($\Delta \gg \Omega $), the drive Hamiltonian can be approximated as a time-independent interaction $H_d^\mathrm{eff}={\Omega^2}/{(4\Delta)}\,a^\dagger a$ with the qubit remaining in the ground state, resulting in a drive-induced frequency shift ${\Omega^2}/{(4\Delta)}$ to the cavity frequency $\omega_c$. As a demonstration in our experiment, the cavity $|0\rangle \rightarrow |1\rangle$ transition frequency is measured and plotted as a function of $\Delta$ in Fig.~\ref{fig4}(b), which agrees well with the theoretical prediction.

Based on the cavity frequency tuning within a bandwidth of about 100~kHz, we conduct independent DP search experiments at each frequency bin using the vacuum state as the initial probe state, with the experimental sequence shown in Fig.~\ref{fig4}(c). The tuning step between adjacent frequency bins is set to approximately 6~kHz, comparable to the DM linewidth, to ensure that the signal appears in one bin. We first characterize the efficiency and false-positive probability across 16 frequency bins within a 100~kHz bandwidth, with the results shown in Fig.~\ref{fig4}(d). The lower false-positive rates ($10^{-4}$) indicate a stable background noise during frequency tuning.

We then perform the DP search experiments using an integration time $\tau \sim T_1^c$ at each frequency bin to extract the accumulated excitation populations using the hidden Markov analysis. As the DM signal cannot be distinguished from background noise in a single bin, we subtract the latter using its weighted average and variance across different bins~\cite{supplement}. This background correction allows us to tighten the upper limits on the DM signal within the 100~kHz frequency range by removing the noise contribution from the total measurement. Figure~\ref{fig4}(e) shows the comparison with previous DP exclusion results, with our results highlighted in Fig.~\ref{fig4}(f) for both the cat-state-based search scheme and frequency-tuning method. Our results demonstrate an unprecedented sensitivity on the order of $10^{-16}$ for DPs within a bandwidth of 100~kHz near 6.5~GHz.

%\section{Conclusions and outlook}

In conclusion, we present a novel quantum-enhanced approach to search for bosonic DPDM using four-component cat states in a high-quality superconducting microwave cavity. This approach is of great advantage because of the inherent error discrimination of the DM-induced displacements from single-photon-loss errors in the background noise. We experimentally demonstrate an 8.1-fold enhancement in the signal photon rate over vacuum states and set a constraint on the DP kinetic mixing angle $\epsilon < 7.32 \times 10^{-16}$. Further enhancement is anticipated using cat states with mean photon numbers exceeding 100~\cite{Vlastakis2013, milul2023}. Furthermore, we employ a parametric sideband drive to actively tune the cavity frequency, enabling DP search across different frequency bins and subtraction of incoherent background noise. This method achieves a sensitivity on the order of $10^{-16}$ within a 100~kHz bandwidth near 6.5~GHz. Both the compass-state-based DP search with a quantum-enhanced efficiency and the frequency-scanning technique with unprecedented sensitivity constitute the complementary aspects of DP searches. When combining these two techniques and integrating them with frequency-tunable qubits~\cite{Chen:2022quj,zhao2025,kang2025,Braggio2025} and sensor networks that exploit quantum coherence~\cite{Chen:2023swh,Shu:2024nmc}, we could explore a promising scalable route towards achieving both higher sensitivity and enhanced detection efficiency in future DP searches in the microwave regime.

\begin{acknowledgments}
We are grateful to Yifan Chen and Yue Zhao for useful discussions. This work was supported by the Quantum Science and Technology-National Science and Technology Major Project (Grants No.~2024ZD0302300, No.~2021ZD0301703), the Guangdong Basic and Applied Basic Research Foundation (Grant No.~2024B1515020013), the National Natural Science Foundation of China (Grants No.~12450006, No.~12025507, No.~12422416, No.~12274198), the Shenzhen-Hong Kong cooperation zone for technology and innovation (Contract No.~HZQB-KCZYB-2020050).
\end{acknowledgments}

\smallskip
\noindent \textbf{\large{}Data availability}{\large\par}
\noindent The data that support the findings of this paper are openly available~\cite{zheng2026}. Further data and analysis codes involved in this study are available from the corresponding authors upon reasonable request. 

%\bibliographystyle{Zou}
%\bibliography{refs}

%Control: production of eprint (-1) disabled
%

\clearpage
\onecolumngrid
\clearpage

\setcounter{section}{0}
\setcounter{equation}{0}
\setcounter{figure}{0}
\setcounter{table}{0}
\makeatletter
\renewcommand{\theequation}{S\arabic{equation}}
\renewcommand{\thefigure}{S\arabic{figure}}
\renewcommand{\bibnumfmt}[1]{[#1]}
\renewcommand{\citenumfont}[1]{#1}

\begin{center}
	\textbf{\large Supplementary Material for ``Quantum-enhanced dark matter search using cat states"} 
\end{center}

\section{Theoretical concept}

\subsection{Dark matter search scheme using cat states}
A coherent state $|\alpha\rangle$ is an eigenstate of the annihilation operator $a$ with eigenvalue $\alpha$, as expressed by $a |\alpha\rangle = \alpha |\alpha\rangle$. This state can be prepared by applying a displacement operator $D(\alpha) = e^{\alpha a^\dagger - \alpha^* a}$ to the vacuum state, yielding $|\alpha\rangle = D(\alpha) |0\rangle$. The coherent state exhibits a Poisson distribution in the Fock basis, parameterized by $\alpha$:
\begin{equation}
	|\alpha\rangle = e^{-\frac{1}{2}|\alpha|^2} \sum_{n=0}^{\infty} \frac{\alpha^n}{\sqrt{n!}} |n\rangle.
\end{equation}
The product of two displacement operators yields an identity:
$D(\beta)D(\alpha) = e^{(\beta\alpha^* - \beta^*\alpha)/2} D(\alpha + \beta)$, which allows us to compute the inner product of two coherent states: $\langle \beta | \alpha \rangle = e^{-\frac{1}{2}(|\alpha|^2 + |\beta|^2 - 2 \beta^* \alpha)}$.

A two-comoponent Schr\"odinger cat state is a superposition of two coherent states with opposite parameters: $|\phi_\pm\rangle = \mathcal{N}_\pm (|\alpha\rangle \pm |-\alpha\rangle)$, where $\mathcal{N}_\pm = {1}/{\sqrt{2(1 \pm e^{-|\alpha|^2})}} \approx {1}/{\sqrt{2}}$ is the normalization constant. Cat states contain only even (odd) Fock states:
\begin{align}
	|\phi_+\rangle &= 2 \mathcal{N}_+ e^{-\frac{1}{2} |\alpha|^2} \sum_{n=0}^{\infty} \frac{\alpha^{2n}}{\sqrt{(2n)!}} |2n\rangle, \\
	|\phi_-\rangle &= 2 \mathcal{N}_- e^{-\frac{1}{2} |\alpha|^2} \sum_{n=0}^{\infty} \frac{\alpha^{2n+1}}{\sqrt{(2n+1)!}} |2n+1\rangle.
\end{align}
Thus, these states have definite parity: $P |\phi_\pm\rangle = \pm |\phi_\pm\rangle$ with $P = e^{i \pi a^\dagger a}$. Applying the annihilation operator changes the parity of a cat state: $a |\phi_\pm\rangle \propto |\phi_\mp\rangle$. 

A compass state (four-component cat state) is a superposition of four coherent states:
\begin{equation}
	|\phi_0\rangle = \mathcal{N}_{4,0} (|\alpha\rangle + |i\alpha\rangle + |-\alpha\rangle + |-i\alpha\rangle),
\end{equation}
where $\mathcal{N}_{4,0}$ denotes the normalization constant. The other three compass states can be obtained by successive application of the annihilation operator:
\begin{align*}
	a |\phi_0\rangle\propto	|\phi_3\rangle &= \mathcal{N}_{4,3} (|\alpha\rangle + i|i\alpha\rangle - |-\alpha\rangle - i|-i\alpha\rangle), \\
	a |\phi_3\rangle\propto	|\phi_2\rangle &= \mathcal{N}_{4,2} (|\alpha\rangle - |i\alpha\rangle + |-\alpha\rangle - |-i\alpha\rangle), \\
	a |\phi_2\rangle\propto	|\phi_1\rangle &= \mathcal{N}_{4,1} (|\alpha\rangle - i|i\alpha\rangle - |-\alpha\rangle + i|-i\alpha\rangle),
\end{align*}
and finally, applying $a$ to $|\phi_1\rangle$ returns to $|\phi_0\rangle$, such that $a |\phi_1\rangle\propto |\phi_0\rangle$ with all $\mathcal{N}_{4,j} \approx 1/2$ for $\alpha\gg1$.

The photon number distribution of compass states exhibits definite modulo 4 properties:
\begin{equation}
	|\phi_j\rangle = 4 \mathcal{N}_{4,j} e^{-\frac{1}{2}|\alpha|^2} \sum_{n=0}^{\infty} \frac{\alpha^{4n+j}}{\sqrt{(4n+j)!}} |4n+j\rangle,
\end{equation}
thereby identifying them as eigenstates of the operator $P_4 = e^{i \frac{\pi}{2} a^\dagger a}$ with $P_4 |\phi_j\rangle = e^{ij \pi / 2} |\phi_j\rangle$.

In general, an $M$-component cat state is a superposition of $M$ coherent states with parameters equidistantly placed on a circle of radius $|\alpha|$:
\begin{equation}
	|\phi_{M,j}\rangle = \mathcal{N}_{M,j} \sum_{k=0}^{M-1} e^{-ij \varphi_k} |e^{i \varphi_k} \alpha\rangle,
\end{equation}
where $\mathcal{N}_{M,j} \approx M^{-1/2}$, and $\varphi_k = 2\pi k / M$. These states are eigenstates of the operator $P_M = e^{i \frac{2\pi}{M} a^\dagger a}$. In the Fock basis, they are expressed as:
\begin{equation}
	|\phi_{M,j}\rangle = M \mathcal{N}_{M,j} e^{-\frac{1}{2} |\alpha|^2} \sum_{n=0}^{\infty} \frac{\alpha^{Mn+j}}{\sqrt{(Mn+j)!}} |Mn+j\rangle.
\end{equation}

The time evolution of the aforementioned states, in the zero-temperature limit, can be described by the master equation in the standard Lindblad form:
\begin{equation}\label{meq}
	\frac{d\rho}{dt} = -\frac{i}{\hbar}[H,\rho] + \frac{\kappa}{2}[2a\rho a^\dagger - a^\dagger a \rho - \rho a^\dagger a],
\end{equation}
where $\kappa$ denotes the single-photon-loss rate, equivalent to the inverse of the single-photon lifetime: $\kappa = {1}/{T_1^c} = {\omega}/{Q}$. Solving the master equation reveals that the amplitude of a coherent state decays as follows: $|\alpha\rangle \rightarrow |\alpha(t)\rangle = |\alpha e^{-\kappa t}\rangle$.

For cat states, the density matrix of the $j$-th cat state $\rho_{M,j} = |\phi_{M,j}\rangle \langle \phi_{M,j}|$ during the time evolution is given by:
\begin{equation}
	\rho_{M,j}(t) = \mathcal{N}_{M,j}^2 \sum_{k,l} e^{-ij(\varphi_k - \varphi_l) - |\alpha|^2(1 - e^{-\kappa t})(1 - e^{i(\varphi_k - \varphi_l)})} |e^{i\varphi_k - \kappa t/2} \alpha\rangle \langle e^{i\varphi_l - \kappa t/2} \alpha|.
\end{equation}
The transition probability between different cat states is expressed as:
\begin{align}
	P_{j \rightarrow l}(t) = \text{Tr}[\rho_j(t)\rho_l(0)] = M \mathcal{N}_{M,j}^2 \mathcal{N}_{M,l}^2 \sum_k e^{-i(j-l)\varphi_k - |\alpha|^2(1 - e^{-\kappa t})(1 - e^{i\varphi_k})}.
	\label{eq:tp}
\end{align}

Taking $j = l$ and expanding to the first order in time, we obtain: $P_{jj}(t) \approx 1 - |\alpha|^2 \kappa t$, which indicates that the cat state has a reduced lifetime of $\tau_\alpha = {T_1^c}/{|\alpha|^2}$.

It is convenient to use the Wigner function $W(z) = {2} \text{Tr}[P D^\dagger(z) \rho D(z)]/{\pi}$ to visualize coherent states and their superpositions. For a state displaced by $\beta$, the density matrix undergoes a transformation: $\rho \rightarrow D(\beta) \rho D^\dagger(\beta)$, and the corresponding Wigner function shifts by $\beta$ with $W'(z) = W(z - \beta)$ in phase space

\begin{figure}[b]
	\centering
	\includegraphics{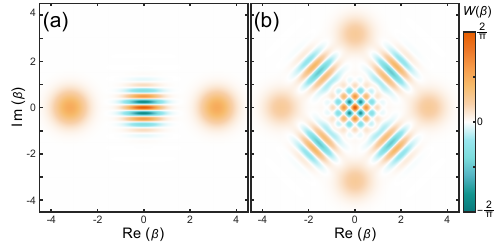}
	\caption{Wigner functions of (a) cat state, and (b) compass state for $\alpha=\sqrt{10}$.}
	\label{fig:SM01}
\end{figure}

The Wigner function for the vacuum state $\ket{0}$ is a standard Gaussian wave packet centered at the origin of the complex plane with $W(z) = \frac{2}{\pi} e^{-2|z|^2}$. For a coherent state $\ket{\alpha} = D(\alpha) \ket{0}$, the Wigner function is also a Gaussian wave packet, but centered at $z = \alpha$ with $W(z) = \frac{2}{\pi} e^{-2|z - \alpha|^2}$. Thus, the sensitivity for measuring the amplitude of weak displacements using the vacuum state (or coherent states) is limited by the standard quantum limit (SQL) due to the unit Gaussian width of the wave packet.

Now, consider the cat states $\ket{\phi_\pm(\alpha)}$ with $\alpha$ real, positive, and large enough to approximate the normalization factors as $1/\sqrt{2}$. In this limit, the Wigner function is given by:
\begin{equation}
	W(z) = \frac{2}{\pi}  \left\{ e^{-2|z-\alpha|^2}+e^{-2|z+\alpha|^2} \pm 2e^{-2|z|^2}\cos[4\alpha \text{Im} z] \right\},
\end{equation}
which comprises two Gaussian wave packets centered at $z = \pm \alpha$, accompanied by an interference term between them, as shown in Fig.~\ref{fig:SM01}(a). This interference fringe is an important nonclassical resource for improving the measurement sensitivity of small displacements along the imaginary axis.

Here, we compare the signal strength for the same displacement $\beta$ between the vacuum state, a single-photon Fock state, and a cat state with opposite parity: 
\begin{align} |\langle{1| D(\beta)|0}\rangle|^2 &\approx \left| \langle 1|\beta a^\dagger|0 \rangle \right|^2 = |\beta|^2, 
	\\
	|\langle{\phi_\mp| D(\beta)|\phi_\pm}\rangle|^2 &\approx \left| \langle \phi_\mp|\beta a^\dagger - \beta^* a|\phi_\pm \rangle \right|^2 = 4|\alpha|^2 (\text{Im} \beta)^2. 
\end{align} 
The response to the imaginary component of the displacement is enhanced by an additional factor of $4|\alpha|^2$, facilitating signal detection.

A limitation of employing cat states is that the enhancement is restricted to the imaginary part of $\beta$, and decoherence effects may eventually compromise the signal, as both the dark matter (DM) signal and background noise can alter the parity. These challenges can be mitigated by using a compass state. As shown in Fig.~\ref{fig:SM01}(b), the Wigner function of a compass state consists of four Gaussian wave packets centered at $z=i^k\alpha$, with interference fringes emerging in both directions at the center for small $\beta$. Using $|\phi_0\rangle$ as the initial state and $|\phi_1\rangle$ as the signal indicator, the signal probability can be expressed as:
\begin{equation}
	\left| \langle \phi_1| D(\beta)|\phi_0 \rangle \right|^2 \approx \left| \langle \phi_1|\beta a^\dagger|\phi_0 \rangle \right|^2 = |\alpha|^2 |\beta|^2,
\end{equation}
which exhibits an enhancement factor of $|\alpha|^2$ for displacement $\beta$ with an arbitrary phase. The decoherence effect primarily induces transitions from $|\phi_0\rangle$ to $|\phi_3\rangle$, rather than affecting $|\phi_1\rangle$, as is evident from the transition probability from $|\phi_0\rangle$ to $|\phi_1\rangle$:
\begin{equation}
	P_{0 \rightarrow 1}(t) = \text{Tr}[\rho_0(t) \cdot \rho_1(0)] = \frac{1}{6} \left( |\alpha|^2 \kappa t \right)^3 + O\left( t^4 \right),
\end{equation}
which is two orders of magnitude smaller than $P_{0 \rightarrow 3}\approx |\alpha|^2 \kappa t $ for small times ($ t \ll 1/|\alpha|^2 \kappa$).

\subsection{Dark matter signal}
In the interaction basis, the dark photon (DP) $A'^\mu$ with mass $m_{A'}$ couples to the Standard Model photon $A^\mu$ through the Lagrangian
\begin{equation}
	\mathcal{L}_{A'}\supset \epsilon m^2_{A'} A^{\prime \mu} A_{\mu},
\end{equation}
which defines an effective current $J_{\text{eff}}^{A^\prime\,\mu} = \epsilon m^2_{A'} A^{\prime \mu}$.
Here, $\epsilon$ denotes the kinetic mixing coefficient. 

The electromagnetic fields within a cavity of volume $V$ are quantized as bound states. In the Coulomb gauge, the vector potential is parameterized as 
\begin{equation}
	\vec{A}  = \sum_n \frac{1}{\sqrt{2\omega_{c,n}}} {a}_n^\dagger \vec{\epsilon}_n(\vec{x}\,)e^{i\omega_{c,n}\, t}+ h.c.,
	\label{eq:Aq}
\end{equation}
where the summation extends over different modes labeled by $n$. Here, ${a}_n$ (${a}_n^\dagger$) represents the annihilation (creation) operator for a mode characterized by eigenfrequency $\omega_{c,n}$ and wave function $\vec{\epsilon}_n(\vec{r})$. The wave functions must satisfy the orthogonality condition: $\int_\text{V}\vec{\epsilon}_m^{\ *} \, \vec{\epsilon}_n\,  dV  = \delta_{mn}$. For the analysis focusing on the $\rm{TM}_{010}$ mode of the cylindrical cavity considered in the main text, the mode label $n$ is omitted in the subsequent discussions. 

To characterize the cavity signal induced by the DM field, one can either model the DM field by a density matrix diagonal in momentum space~\cite{Derevianko:2016vpm}, or consider it as a superposition of $N\gg 1$ classical freely propagating plane waves \cite{Foster:2017hbq}
\begin{equation}
	\vec{A'}=\sum_{i=1}^N \vec{A'}_i(t,\vec{x})=\sum_{i=1}^N {A'}_{i,0}\vec{\epsilon}_i\cos(\omega_i t-\vec{k}_i\cdot\vec{x}+\varphi_i)
\end{equation}
where ${A'}_{i,0}$ denotes the effective magnitude of gauge potential vector, uniform across all particles and related to the local DM density in the galaxy by $\rho_{\mathrm{DM}}={N}m_{\mathrm{DM}}^2{A'}_{i,0}^2/2=0.4 \text{ GeV/cm}^3$. Here, $m_{\mathrm{DM}}=m_{A'}$ represents the mass of the DP particle, $\vec{\epsilon}_i$ is the isotropically distributed polarization vector, $\omega_i=m_{\mathrm{DM}}(1+v_i^{2}/2)$ is the Compton frequency, $\vec{k_i}=m_{\mathrm{DM}}\vec v_i$ is the galactic wave vector associated with the apparatus motion through the DM halo, and $\varphi_i$ is a random phase distributed uniformly from 0 to $2\pi$. The speed of the DM field follows the Maxwellian distribution according to the standard halo model of cold DM
\begin{equation}
	f_{\mathrm{DM}}(v)=\frac{v}{\sqrt{\pi}v_{vir} v_g}e^{-(v+v_g)^2/v_{vir}^2}(e^{4vv_g/v_{vir}^2}-1),
\end{equation}
where $v_{vir}\approx 220$~km/s is the virial velocity and $v_g\approx 232$ km/s is the speed of the Sun relative to the halo rest frame. The energy distribution is then given by
\begin{equation}
	f_{\mathrm{DM}}(\omega_{\mathrm{DM}})=\frac{1}{ \sqrt{2({\omega_{\mathrm{DM}}}{m_{\mathrm{DM}}}-m_{\mathrm{DM}}^2)}}f_{\mathrm{DM}}\left[\sqrt{2\left(\frac{\omega_{\mathrm{DM}}}{m_{\mathrm{DM}}}-1\right)}\right].
\end{equation}

Note that the DM line profile is strictly zero for $\omega_{\mathrm{DM}}< m_{\mathrm{DM}}$, and the maximum $f_{\mathrm{DM}}(\omega_m)\approx 10^6/m_{\mathrm{DM}}$ occurs at $\omega_m=m_{\mathrm{DM}}(1+v_m^2/2)\approx (1+3\times 10^{-7})m_{\mathrm{DM}}$ for $v_{m}\approx 237$~km/s.

The interaction Hamiltonian between the DP background and the cavity is given by:
\begin{align}
	H_{\mathrm{int}} &=\int_\text{V}\,\vec{A}\cdot \vec{J}_{\rm eff}\,dV=\epsilon m^2_{\mathrm{DM}} \sum_i\int_\text{V}\,\vec{A}\cdot \vec{A}^\prime_i \,dV\\
	&
	= \frac{\epsilon m^2_{\mathrm{DM}}}{\sqrt{2 \omega_c}}\sum_i\, A'_{i,0}\sqrt{V_{\rm eff,i}} \, {a}^{\dagger}\, e^{i(\omega_c-\omega_i)\, t-i\varphi_i} + h.c..\label{eq:cavalp}
\end{align}
where the rotating wave approximation is applied and $V_{\rm eff,i}$ is the effective volume defined as
\begin{equation}
	V_{\rm eff,i} = \left|\int_V \vec{\epsilon} \cdot \vec{\epsilon}_i e^{-i\vec{k}_i\cdot\vec{x}}dV\right|^2.
	\label{eq:etan}
\end{equation}

In the presence of a DP field \( \vec{A'}_i(t,\vec{x})= A'_{i,0}\vec{\epsilon}_i\cos(\omega_i t-\vec{k}_i\cdot\vec{x}+\varphi_i) \), the Hamiltonian can be interpreted as a coherent drive with detuning \( \delta_i=\omega_i-\omega_c \) and amplitude \( \Omega_i = \epsilon m^2_{\mathrm{DM}}A'_{i,0}\sqrt{V_{\rm eff,i}/{2\omega_c}} \).

Under resonance condition \( \delta_i=0 \), the DM field acts as a displacement operator, expressed as \( e^{-iH_i\tau}= D(\beta) \) with $\beta=-i\Omega_i e^{i\varphi_i}\tau$. In the off-resonant case, the time evolution of the cavity state is given by
\begin{equation}
	|\psi_i(t)\rangle= D(z_i) e^{-i\int_0^tK_i(\tau)d\tau}e^{-ia^\dagger 
		a\omega_c t}|\psi(0)\rangle ,
\end{equation}
where
\begin{align}
	z_i(t)&=-\Omega_i e^{-i\omega_c t-i\varphi_i}({1-e^{-i\delta_i t}})/{\delta_i},\\
	K_i(t)&=\Omega_i^2 ({1-\cos{\delta_i t}})/{\delta_i}.
\end{align}

For a cavity initialized in the ground state \( |\psi(0)\rangle=|0\rangle \) and assuming \( \Omega_i\ll 1 \), the state evolves to $|\psi_i(t)\rangle\approx c_{0,i}|0\rangle+c_{1,i}|1\rangle $, where \( c_{0,i}\approx 1 \) and $c_{1,i}\approx e^{-i(\omega_i+\omega_c)t/2 -i\varphi_i}\Omega_i  \frac{\sin(\delta_it/2)}{\delta_i/2}$. 
The total effect of the DP background drives the cavity from the vacuum state to $|\psi(t)\rangle\approx c_0|0\rangle+c_1 |1\rangle $ with \( c_0\approx 1 \) and \( c_1=\sum_i c_{1,i} \).

The ensemble average of an observable \( \mathcal{O} \) is defined as
\begin{equation}
	\langle\mathcal{O}\rangle=\prod_{i,j,k}\int d^3\vec{v}_if_{\mathrm{DM}}(\vec v_i)\int_0^{2\pi}\frac{d\varphi_j}{2\pi}\int \frac{d^2\hat{\epsilon}_k}{4\pi} \mathcal{O}.
\end{equation}

Then the effective volume of a cylindrical cavity can be estimated as $V_{\rm eff}=\langle V_{\rm eff,i}\rangle\approx 0.23 V$ and the excitation probability of the cavity state can be calculated as
\begin{equation}\label{eq:p1t}
	p_1(t)=\langle|c_1|^2\rangle\equiv  N\Omega^2 g(t),
\end{equation}
where \( N\Omega^2=\epsilon^2 m^2_{\mathrm{DM}}\rho_{\mathrm{DM}}V_{\rm eff}/{\omega_c} \) and
\begin{equation}\label{eq:gt}
	g(t)=\int d\omega f_{\mathrm{DM}}(\omega)\left(\frac{\sin((\omega-\omega_c) t/2)}{(\omega-\omega_c)/2}\right)^2.
\end{equation}

Here, we take $\omega_c=\omega_m=(1+3\times 10^{-7})m_{\mathrm{DM}}$, where $f_{\mathrm{DM}}$ reaches its maximum $f_{\mathrm{DM}}(\omega_m)\approx 0.98\times10^6/m_{\mathrm{DM}}$.
When \( (\omega-\omega_m) t\ll 1 \), we can approximate $\sin((\omega-\omega_m) t/2) \approx (\omega-\omega_m) t/2$, leading to \( g(t)\approx t^2 \), which indicates coherent signal buildup within the DM coherence time. 

\begin{figure}[htbp]
	\centering
	\includegraphics{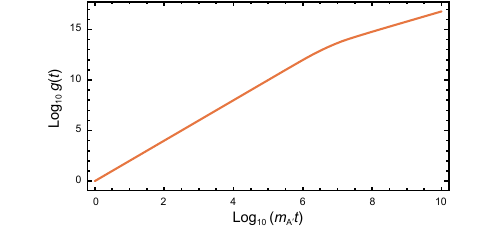}
	\caption{Plot of function \( g(t) \) showing piecewise behavior.}
	\label{fig:gt}
\end{figure}

Conversely, for sufficiently long integration time, we can approximate
$\left(\frac{\sin((\omega-\omega_m) t/2)}{(\omega-\omega_m)/2}\right)^2 \approx 2\pi t \delta(\omega-\omega_m)$,
resulting in \( g(t)\propto \tau_{\mathrm{DM}}t \), which indicates coherence loss due to the dispersion caused by variations in the velocities of its constituent DM particles. 
The DM coherence time is then given by
$$\tau_{\mathrm{DM}}=2\pi f_{\mathrm{DM}}(\omega_m) \approx \frac{2\pi\times10^6}{m_{\mathrm{DM}}}$$ 

As depicted in Fig.~\ref{fig:gt}, the function \( g(t) \) exhibits a piecewise behavior, with a transition occurring around \( \tau_{\mathrm{DM}}\):
\begin{equation}
	g(t)\simeq\begin{cases}
		t^2  & \text{if } 0 < t < \tau_{\mathrm{DM}},\\
		\tau_{\mathrm{DM}}t & \text{if } t > \tau_{\mathrm{DM}}.
	\end{cases}
\end{equation}

Similarly, when initializing the cavity with the compass state $\ket{\phi_0(\alpha)}$, we can obtain that the state transition probability to $\ket{\phi_1(\alpha)}$ is given by
\begin{equation}\label{eq:agt}
	p_{0\rightarrow1}(t)=N\Omega^2|\alpha|^2 g(t),
\end{equation}
which exhibits an enhancement factor of $|\alpha|^2$ compared to Eq.~\eqref{eq:p1t}.

\section{Experimental details}
\subsection{Experimental setup}
Our experiment utilized a circuit quantum electrodynamics~\cite{blais2021} device, which consists of a three-dimensional (3D) niobium cylindrical cavity~\cite{reagor2013} whose fundamental mode serves as the storage mode, a superconducting transmon qubit~\cite{koch2007}, and a Purcell-filtered stripline readout resonator~\cite{axline2016}.  

\begin{figure}[b]
	\centering
	\includegraphics[width=1\textwidth]{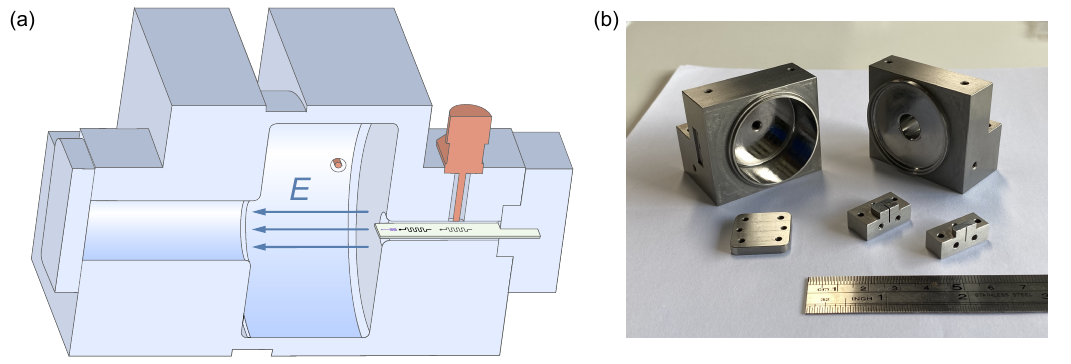}
	\caption{The 3D model and machined pieces of the cylindrical cavity. (a) The sections of the cylindrical cavity showing the inner structure of the cavity and the location of the assembled chip. The arrows represent the electrical fields of the cavity TM$_{010}$ mode. (b) The picture of the machined cavity pieces before e-beam welding.}
	\label{fig:cavity}
\end{figure}

\begin{figure}
	\centering
	\includegraphics{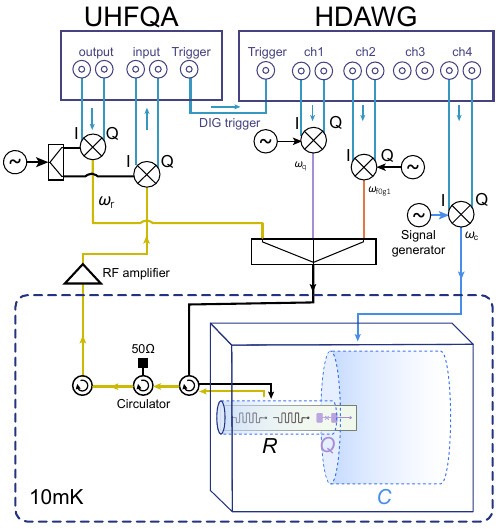}
	\caption{The wiring diagram of the experimental circuitry and device.}
	\label{fig:setup}
\end{figure}

The 3D cavity, designed with a cylindrical shape, aims to maximize the conversion rate of DPs into resonating real photons. The cavity's inner space is illustrated in Fig.~\ref{fig:cavity}(a), where the main cylinder cavity with a diameter of 36~mm and a height of 19~mm has a total volume of 19.34~cm$^3$. There are two coaxial cylindrical tunnels: the larger (10~mm diameter) for injecting the chemical polishing solution, and the narrower (3.2~mm diameter) for housing a sapphire chip patterned with a qubit and a readout resonator. Finite element analysis indicates that the resonance frequency of the fundamental TM$_{010}$ mode is around 6.5 GHz with its electric field primarily aligning with the cylinder's axis, maximizing the cavity's form factor, as depicted in Fig.~\ref{fig:cavity}(a). The cavity was first precision-machined from high-purity niobium (RRR$\geqslant$300) into two halves, then joined and e-beam welded together. Here, RRR (Residual Resistivity Ratio) quantifies the purity and quality of niobium. A photo of the cavity halves before welding is shown in Fig.~\ref{fig:cavity}(b). The typical axial shrinkage after e-beam welding is about 0.50 $\pm$ 0.15 mm. Surface treatments such as chemical polishing and annealing enhanced the surface quality, thereby improving the cavity coherence time.

The superconducting transmon qubit and the Purcell-filtered readout resonator are patterned on the sapphire chip, which is inserted into the narrower tunnel to the cylindrical cavity. The qubit is designed with two antenna pads: one for coupling to the 3D cavity for storage-mode manipulation and the other for coupling to the readout resonator for qubit-state readout. The Josephson junction of the qubit is an Al-Al$_2$O$_x$-Al trilayer tunnel junction, created via the double-angle evaporation technique. Both the qubit antenna pads and the readout striplines are made of tantalum to enhance the qubit coherence times~\cite{place2021,wang2022}. The Purcell-filtered readout resonator, patterned as two meandering metal strips on the sapphire chip and integrated with the chip-housing tunnel's conducting wall, forms two half-wavelength transmission-line resonators. 

The 3D cavity and the sapphire chip were assembled together and then mounted on the 10~mK cold plate of a dilution refrigerator. Commercial microwave electronics, including a high-density arbitrary waveform generator (HDAWG) and an ultra-high-frequency quantum analyzer (UHFQA) from Zurich Instruments, are used to control and measure quantum states of the experimental sample. The wiring diagram of the measurement setup is shown in Fig.~\ref{fig:setup}.

\begin{table}[b]
	\centering
	\caption{Measured device parameters. The subscripts $c$, $q$, and $r$ represent the 3D storage cavity, transmon qubit, and readout resonator, respectively.}
	\begin{tabular}{cccc}
		\hline
		\hline
		&\\[-0.9em]
		Measured parameters & 3D storage cavity & Transmon qubit & Readout resonator \\\hline
		&\\[-0.9em]
		Mode frequency $\omega_{c,q,r}/2\pi$ & 6.442 GHz& 5.205 GHz& 7.950 GHz\\
		&\\[-0.9em]
		Self-Kerr nonlinearity $K_{c,q,r}/2\pi$ & 498 Hz& 235.1 MHz& -\\
		&\\[-0.9em]
		Mode dispersive shift $(\chi_{qc},\chi_{qr})/2\pi$ & \multicolumn{3}{c}{\begin{tabular}{cc} 0.6 MHz\quad\quad & \quad\quad 3.4 MHz\end{tabular}}\\
		&\\[-0.9em]
		\hline
		&\\[-0.9em]
		Relaxation $T_{1}$ & 4.6 ms& 175.3 $\upmu$s& 188 ns\\
		&\\[-0.9em]
		Ramsey coherence $T_{2}$ & 4.5 ms& 119.4 $\upmu$s& -\\
		&\\[-0.9em]
		Thermal population $\bar{n}_{c,q}$ & 0.01\%& 1.3\%& - \\
		&\\[-0.9em]
		\hline
		\hline
	\end{tabular}
	\label{tab:H-param}
\end{table}

\subsection{System Hamiltonian}
The full system Hamiltonian up to the second order can be written as
\begin{equation}
	H= \sum_p\left(\omega_p a_p^\dagger a_p - \frac{1}{2}\chi_{pp}a_p^\dagger a_p a_p^\dagger a_p \right) - \sum_{p\neq p'}\chi_{pp'}a_p^\dagger a_p a_{p'}^\dagger a_{p'}
	\label{eq:Hamiltonian}
\end{equation}
where the indices \( p \) and \( p' \) correspond to \( c, q, \) and \( r \), representing the 3D cavity, transmon qubit, and readout resonator modes, respectively, with corresponding frequencies \( \omega_p \). Here, \( a_p \) is the annihilation operator, \( \chi_{pp} = K_p \) represents the self-Kerr nonlinearity of mode \( p \), and \( \chi_{pp'} \) denotes the cross-Kerr interaction (mode dispersive shift) between modes \( p \) and \( p' \). Higher-order terms are neglected as they are expected to be small. The values of these parameters are experimentally calibrated and summarized in Table~\ref{tab:H-param}.

Since the readout cavity remains in its ground state except during qubit readout, its contribution to the interaction between the storage cavity and the transmon is neglected. By truncating the transmon qubit into its lowest two energy levels and neglecting the sub-leading self-Kerr interactions, we can obtain a simplified dispersive Hamiltonian:
\begin{equation}\label{HD}
	H = \omega_c a^\dagger a + \omega_q|e\rangle\langle e| - \chi a^\dagger a |e\rangle\langle e|
\end{equation}
involving the qubit and the storage cavity, as described in the main text. Here, the operator $a$ and the coefficient $\chi$ are short for $a_c$ and $\chi_{qc}$ in Eq.~\eqref{eq:Hamiltonian}, respectively.

\section{Characterization and analysis}
\subsection{Parity measurement calibration}
Parity measurements are essential for preparing and characterizing cat states, as well as for repeated quantum non-demolition (QND) measurements of the final cavity state after integrating the DM signal. This process employs the Ramsey interferometry method, which includes a qubit $\pi/2$ pulse, a delay time $t$, another qubit $-\pi/2$ pulse, and a final qubit state projection. The two $\pi/2$ pulses act as beam-splitters for the qubit states, while the free evolution during the delay time entangles the qubit and cavity states. This process corresponds to a sinusoidal photon number filter~\cite{Deng2024} operation $P_S(\theta)\approx \sum_n{\cos{\frac{n\theta}{2}}|n\rangle\langle n|}$, which transforms arbitrary input cavity state $|\psi_\mathrm{in}\rangle$ to output state $|\psi_\mathrm{out}\rangle \propto P_S(\theta)|\psi_\mathrm{in}\rangle$ with $\theta=\chi t_p$. The delay time $t_p$ is set to $\pi/\chi$ for mod-2 parity measurements and $\pi/2\chi$ for mod-4 parity measurements, corresponding to $P_S(\pi)$ and $P_S(\pi/2)$, respectively. These delay times are experimentally optimized by performing parity measurements on a known cavity state with a certain parity.

\begin{table}[b]
	\centering
	\caption{Demolition probability $p_d$ and its standard deviation $\sigma_{p_d}$ of the readout and the parity measurements.}
	\begin{tabular}{p{2in}p{1in}p{1in}}
		\hline
		\hline
		Measurement mode & $p_d$  & $\sigma_{p_d}$ \\\hline
		&\\[-0.9em]
		Only readout  & 0.8\% &0.4\%\\
		&\\[-0.9em]
		Parity measurement (mod 2)  & 1.2\% & 0.3\%\\
		&\\[-0.9em]
		Parity measurement (mod 4)  & 1.3\% &0.4\% \\
		&\\[-0.9em]
		\hline
		\hline
	\end{tabular}
	\label{tab:QND}
\end{table}

Ideal parity measurements retrieve only the parity information without disturbing the cavity states. However, in practice, these measurements may introduce additional relaxation errors in the cavity states. The QND nature of the parity measurements can be assessed by measuring the cavity coherence times with interleaved repeated parity measurements~\cite{Sun2014}. The decoherence of cat states is a combination of the bare energy relaxation $\kappa_s$ and a demolition probability $p_d$. The demolition probability can be determined by fitting the measured coherence times against the repetition interval between parity measurements, as listed in Table~\ref{tab:QND}. The table also lists the demolition probability of the readout operation, obtained by interleaving only qubit state projections.

\subsection{Cavity drive pulse calibration}
A coherent drive pulse applied to the cavity generates a coherent state $|\beta\rangle$, which is useful for both cat-state preparation and DM detector calibration. Calibrating the drive pulse is essential for determining the injected photon number $|\beta|^2$.

\begin{figure}
	\centering
	\includegraphics[width=1\textwidth]{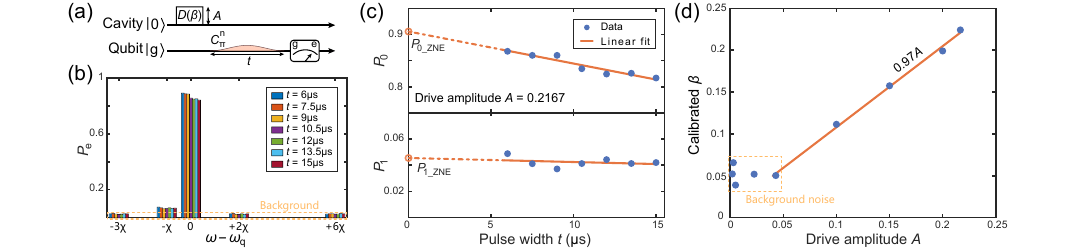}
	\caption{Calibration of the cavity displacement drive. (a) The experimental sequence of the calibration. A cavity displacement drive with an amplitude $A$ is applied to generate a coherent state $|\beta\rangle$, followed by a photon-number-selective $\pi$ pulse and a qubit readout to extract injected photon numbers. (b) Measured qubit excited state probabilities as a function of the frequency detuning $\omega-\omega_q$ of the selective drive for various selective pulse durations. The dashed horizontal line indicates the qubit measurement background. (c) Extracted photon number populations $P_0$ and $P_1$ as a function of the duration of the selective $\pi$ pulse. The orange line is a linear fit, with the $y$-intercept representing corrected photon-number populations mitigated by qubit relaxation errors during the selective pulse. (d) The calibrated $|\beta|$ as a function of the nominal amplitude $A$ of the coherent drive pulse. The orange line is a linear fit to the linear portion of the plot. }
	\label{fig:cavity-drive}
\end{figure}

The calibration procedure is illustrated in Fig.~\ref{fig:cavity-drive}(a). The cavity is first displaced to a coherent state $|\beta\rangle$ by a drive pulse with a fixed duration and nominal amplitude $A$. The qubit spectrum is then measured using the photon-number-selective $\pi$ pulse followed by qubit readout. The measured qubit excited probability $P_e$ at the detuned frequency $\omega-\omega_q=-n\chi$ corresponds to the photon number population $P_n$, which follows a Poisson distribution:
\begin{equation}
	P_n=|\langle n|\beta\rangle|^2=\mathrm{e}^{-|\beta|^2}\frac{|\beta|^{2n}}{n!}.
\end{equation}
The photon number $|\beta|^2$ is inferred from the ratio of the amplitudes at detuned frequency of 0 and $-\chi$, corresponding to $P_1/P_0$.

This method becomes challenging for weak drives due to the noisy background in the qubit spectrum. The low population of the Fock-state component $|1\rangle$ in the cavity state, due to weak drives, leads to a low signal-to-noise ratio in the spectrum. The spectrum background is assumed to have at least two contributions: a constant component from measurement errors and a component related to qubit relaxation during the selective $\pi$ pulse.

Measurement error is assessed using readout signals at frequencies far from those corresponding to possible excited photon number components, such as at detuned frequencies $-3\chi$, $+2\chi$, and $+6\chi$. Qubit relaxation error during the selective $\pi$ pulse is mitigated using the zero-noise-extrapolating (ZNE) method~\cite{kandala2019,Kim2023,Cai2023}, where multiple spectra with varying selective pulse durations are acquired. The measured photon-number populations $P_0$ and $P_1$ increase gradually as the duration is shortened, with the $y$-intercept of the linear fit representing the corrected photon-number populations. The experimental results of these calibrations are presented in Fig.~\ref{fig:cavity-drive}(b) and Fig.~\ref{fig:cavity-drive}(c).

After subtracting the two error contributions, the corrected $P_1/P_0$ ratio yields the injected photon number $|\beta|^2$. By repeating the aforementioned experiments while varying the nominal drive amplitude $A$, we extract the calibrated $|\beta|$ as a function of $A$, which is shown in Fig.~\ref{fig:cavity-drive}(d). The experimental results indicate a linear relationship between the calibrated $\beta$ and the drive amplitude $A$ for $A>0.05$, which is used to calibrate the injected photon numbers. The measured background noise at small drive amplitude may be attributed to the cavity thermal population and remaining measurement-induced errors.

\begin{figure}[b]
	\centering
	\includegraphics[width=1\textwidth]{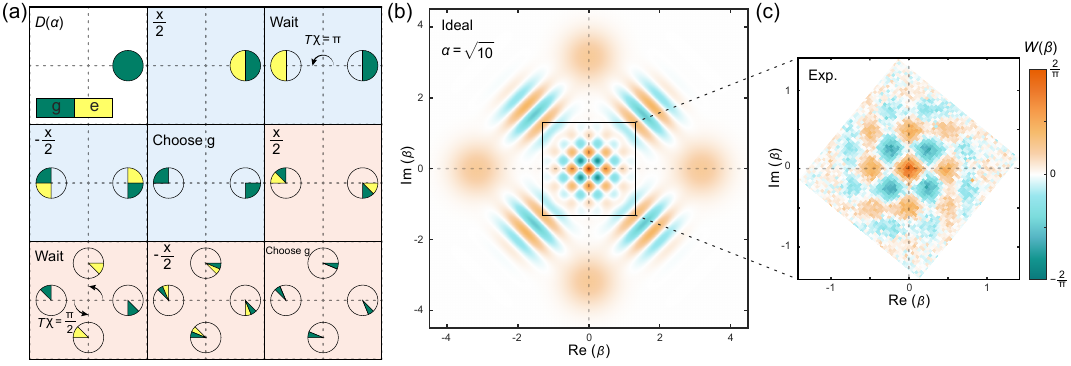}
	\caption{Preparation and measurement of cat states. (a) Illustration of the phase-space evolution for generating a four-component cat state. The circles represent the coherent-state components, and the green (yellow) stands for the portion that is entangling with the qubit ground (excited) state. (b) The ideal Wigner function of the compass state $|\phi_0(\alpha)\rangle$ with $\alpha=\sqrt{10}$. (c) The measured Wigner function of $|\phi_0\rangle$ with $\alpha=\sqrt{10}$, which is manually rotated into the correct rotating frame for comparison with the ideal one.}
	\label{fig:cats}
\end{figure}

\subsection{Preparation and measurement of cat states}
The generation and characterization of cat states in the 3D cavity are performed using the Ramsey interferometry methods, as described in Fig.~\ref{fig:cats}(a).

When the qubit initialized in the ground state \( |g\rangle \) undergoes a \( {\pi}/{2} \) pulse about the $x$-axis, it transforms into the superposition state \( (|g\rangle +i |e\rangle)/\sqrt{2} \). The cavity-qubit dispersive interaction, given by \( H_{\text{int}} = -\chi a^\dagger a |e\rangle\langle e| \), introduces a photon-number-dependent phase shift.

After a wait time \( t = \pi/\chi \), the superposition state accumulates a relative \( \pi \) phase if the cavity contains an odd number of photons. A subsequent\( -\pi/2 \) pulse about the $x$-axis flips the qubit to \( |e\rangle \). Conversely, if the cavity has an even number of photons, no phase difference accumulates, leaving the qubit in \( |g\rangle \). As a result, the initial cavity-qubit state \( |\alpha\rangle \otimes |g\rangle \) evolves into an entangled state:
\begin{equation}
	|\alpha\rangle\otimes|g\rangle \rightarrow |\phi_+\rangle\otimes|g\rangle + i|\phi_-\rangle\otimes|e\rangle.
\end{equation}
A final qubit readout collapses the qubit in \( |g\rangle \) or \( |e\rangle \), consequently projecting the cavity onto the corresponding cat state \( |\phi_+\rangle \) or \( |\phi_-\rangle \).

To generate a compass state from the prepared two-component cat state, we can repeat the aforementioned procedure while adjusting the wait time between the two \( \pi/2 \) pulses to \( t = \pi/2\chi \). This allows the cavity-qubit system to evolve as:
\begin{align*}
	|\phi_+\rangle\otimes|g\rangle & \rightarrow|\phi_0\rangle\otimes|g\rangle + i|\phi_2\rangle\otimes|e\rangle,\\
	|\phi_-\rangle\otimes|e\rangle & \rightarrow|\phi_1\rangle\otimes|e\rangle + |\phi_3\rangle\otimes|g\rangle.
\end{align*}
A subsequent qubit readout then projects the cavity onto one of the four compass states. For instance, preparing $|\phi_0\rangle$ involves first generating $|\phi_+\rangle$ using a Ramsey sequence with wait time $t=\pi/\chi$ and then projecting onto the target compass state using another Ramsey sequence with wait time $t=\pi/2\chi$. 

Prepared compass states can be examined through photon number population measurements or Wigner tomography, as shown in Fig.~1(d) in the main text and Figs.~\ref{fig:cats}(b-c). The fidelity of these states can be evaluated by comparing the measurement results with those expected for the ideal state.

\begin{figure}[t]
	\centering
	\includegraphics{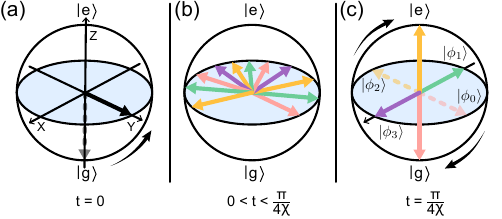}
	\caption{Bloch sphere representations of the qubit states during the Ramsey parity measurement. (a) The first $\pi/2$ pulse rotates the qubit ground state to a superposition state$(\ket{g} +i \ket{e})/\sqrt{2}$. (b) During a wait time of $\pi/2\chi$, the superposition state rotates around the $z$-axis to gain additional phases due to the photon-number-dependent phase shift. (c) At the end of the wait time, the qubit state entangled with the cavity state $\ket{\phi_j}$ accumulates a phase of $j\pi/2$, as depicted by the arrows along the $x$- and $y$-axes. A subsequent $-\pi/2$ rotation about the $x$-axis ($y$-axis) can transform the qubit state back into the $\sigma_z$ subspace, projecting the cavity into the even (odd) cat state. }
	\label{fig:Ramsey}
\end{figure}

In detector characterization and DM search experiments, we first prepare the cavity in the initial state \( \ket{\phi_0} \), and then apply a mimic displacement drive or signal integration of the real DM field, which results in an evolved state. Finally, the transition probability from \( \ket{\phi_0} \) to \( \ket{\phi_1} \) is measured using a similar Ramsey interferometry method with a wait time of $\pi/2\chi$. However, the rotation axis of the second $\pi/2$ pulse should be changed to $y$-axis to distinguish \( \ket{\phi_1} \) from \( \ket{\phi_3} \), as illustrated in Fig.~\ref{fig:Ramsey}. After a single Ramsey parity measurement, the qubit state flips if the cavity is in $\ket{\phi_1}$, remains unchanged if in $\ket{\phi_3}$, and transitions randomly with equal probabilities if the cavity is in $\ket{\phi_0}$ or $\ket{\phi_2}$ states. Repeating this measurement allows discrimination of the evolved cavity states based on three sequence patterns: \textit{alternating} for $\ket{\phi_1}$, \textit{constant} for $\ket{\phi_3}$, and \textit{random} for the even-parity states $\ket{\phi_0}$ or $\ket{\phi_2}$. 

\subsection{Hidden Markov model}

In the ideal scenario, repeated parity measurements on the cavity would alter the states of the cavity and the qubit solely based on their current states.
However, these states are susceptible to errors and may change randomly during free evolution and measurements.
Consequently, observed random sequences may not necessarily indicate that the evolved cavity is in the even-parity states $\ket{\phi_0}$ or $\ket{\phi_2}$.
To account for potential measurement errors and assess the probability of the evolved cavity state, we adopt a three-layer Hidden Markov Model (HMM) approach, as presented in \cite{Dixit2021, Agrawal2024, zhao2025}.

The cavity and qubit states are treated as hidden variables, with their evolution described by a transition matrix $T$ that captures all possible changes. We assume that the cavity state switches between four compass states, $\ket{\phi_j}\in \{\ket{\phi_0}, \ket{\phi_1}, \ket{\phi_2}, \ket{\phi_3}\}$. The transition probability due to single-photon loss from $\ket{\phi_j}$ to $\ket{\phi_{j-1}}$ is $P_{j,j-1}=1-e^{-|\alpha|^2t_m/T_1^c}$, while that due to spontaneous heating from $\ket{\phi_j}$ to $\ket{\phi_{j+1}}$ is $P_{j,j+1}=\bar{n}_c(1-e^{-|\alpha|^2t_m/T_1^c})$. Here, the interval between parity measurements $t_m=1.9\,\mu$s is primarily limited by the readout time ($1.2\,\mu$s) and the time required to actively reset the readout resonator~\cite{McClure2016} to the ground state ($0.3\,\mu$s). The coherence time of the compass state $\ket{\phi_j(\alpha)}$ is approximately as $T_1^c/|\alpha|^2$ with $T_1^c$ representing the cavity single-photon lifetime. $\bar{n}_c$ denotes the steady-state thermal population in the cavity. The high-order multi-photon processes (either relaxation or heating) are neglected, giving $P_{j,j\pm2}=0$. The probability of no errors is then given by $P_{jj}=1-P_{j,j-1}-P_{j,j+1}$.

Next, we assume that the qubit state switches between the ground state \( \ket{g} \) and the first excited state \( \ket{e} \), neglecting higher-order excitations. The qubit can relax, undergo spontaneous heating, or experience dephasing, with probabilities \( P^\downarrow_{eg} = 1 - e^{-t_m/T_1^q} \), $P^\uparrow_{ge}=\bar n_q(1-e^{-t_m/T_1^q})$ and \( P^\phi = 1 - e^{-t_p/T_2^q} \), respectively. Here, $t_p$ is the wait time in the parity measurement. Qubit errors are characterized by \( P_{ge} = P^\uparrow_{ge} + P^\phi \) and \( P_{eg} = P^\downarrow_{eg} + P^\phi \). The no-error probabilities are given by \( P_{gg} = 1 - P_{ge} \) and \(P_{ee} = 1 - P_{eg} \). Here, \( T_1^q \) and \( T_2^q \) represent the qubit's energy relaxation and coherence lifetimes, respectively, and \( \bar{n}_q \) denotes the qubit's steady-state thermal population.

The coherence times and steady-state thermal populations of the qubit and cavity are obtained using standard characterization methods and are listed in Table.~\ref{tab:H-param}. The wait time in the Ramsey interferometry is approximately given by $t_p=\pi/2\chi$, which is further characterized by measuring the parity of the compass state $\ket{\phi_0}$ with varying $t_p$.

Given the joint cavity-qubit hidden states \( S \in \{ \ket{\phi_0 g}, \ket{\phi_0 e}, \ket{\phi_1 g}, \ket{\phi_1 e}, \ket{\phi_2 g}, \ket{\phi_2 e}, \ket{\phi_3 g}, \ket{\phi_3 e} \} \), the transition matrix can be expressed as follows:
\begin{equation}
	\begin{array}{ccc}
		& \begin{matrix}
			|\phi_0g\rangle\ & |\phi_0e\rangle\ & |\phi_1g\rangle\ \ & |\phi_1e\rangle\ \ 
			& |\phi_2g\rangle\ & |\phi_2e\rangle\ & |\phi_3g\rangle\ \ & |\phi_3e\rangle
		\end{matrix} & \\
		T=  & \begin{pmatrix}
			P_{00}/2 & P_{00}/2 & P_{01}P_{ge} & P_{01}P_{gg} & P_{02}/2 & P_{02}/2 & P_{03}P_{gg} & P_{03}P_{ge} \\
			P_{00}/2 & P_{00}/2 & P_{01}P_{ee} & P_{01}P_{eg} & P_{02}/2 & P_{02}/2 & P_{03}P_{eg} & P_{03}P_{ee} \\
			P_{10}/2 & P_{10}/2 & P_{11}P_{ge} & P_{11}P_{gg} & P_{12}/2 & P_{12}/2 & P_{13}P_{gg} & P_{13}P_{ge} \\
			P_{10}/2 & P_{10}/2 & P_{11}P_{ee} & P_{11}P_{eg} & P_{12}/2 & P_{12}/2 & P_{13}P_{eg} & P_{13}P_{ee} \\
			P_{20}/2 & P_{20}/2 & P_{21}P_{ge} & P_{21}P_{gg} & P_{22}/2 & P_{22}/2 & P_{23}P_{gg} & P_{23}P_{ge} \\
			P_{20}/2 & P_{20}/2 & P_{21}P_{ee} & P_{21}P_{eg} & P_{22}/2 & P_{22}/2 & P_{23}P_{eg} & P_{23}P_{ee} \\
			P_{30}/2 & P_{30}/2 & P_{31}P_{ge} & P_{31}P_{gg} & P_{32}/2 & P_{32}/2 & P_{33}P_{gg} & P_{33}P_{ge} \\
			P_{30}/2 & P_{30}/2 & P_{31}P_{ee} & P_{31}P_{eg} & P_{32}/2 & P_{32}/2 & P_{33}P_{eg} & P_{33}P_{ee} 
		\end{pmatrix} & \begin{matrix}
			|\phi_0g\rangle \\ |\phi_0e\rangle \\ |\phi_1g\rangle \\ |\phi_1e\rangle \\
			|\phi_2g\rangle \\ |\phi_2e\rangle \\ |\phi_3g\rangle \\ |\phi_3e\rangle
		\end{matrix}
	\end{array}
\end{equation}

The observations collected in the experiment are sequences of qubit readout signals, \( R \in \{\mathcal{G}, \mathcal{E}\} \), which are governed by the joint hidden states \( S \) and are subject to readout errors \( F_{e\mathcal{G}}\) and \( F_{g\mathcal{E}}\). The no-error probability is given by \( F_{g\mathcal{G}} = 1 - F_{e\mathcal{G}} \) and \( F_{e\mathcal{E}} = 1 - F_{g\mathcal{E}} \). The transition between \( S \) and \( R \) is characterized by the emission matrix:
\begin{equation}
	\begin{array}{ccc}
		& \begin{matrix}
			\mathcal{G}\ & \ \mathcal{E}
		\end{matrix} & \\
		E=  & \begin{pmatrix}
			F_{g\mathcal{G}} & F_{g\mathcal{E}} \\
			F_{e\mathcal{G}} & F_{e\mathcal{E}} \\
			F_{g\mathcal{G}} & F_{g\mathcal{E}} \\
			F_{e\mathcal{G}} & F_{e\mathcal{E}} \\
			F_{g\mathcal{G}} & F_{g\mathcal{E}} \\
			F_{e\mathcal{G}} & F_{e\mathcal{E}} \\
			F_{g\mathcal{G}} & F_{g\mathcal{E}} \\
			F_{e\mathcal{G}} & F_{e\mathcal{E}} 
		\end{pmatrix} & \begin{matrix}
			|\phi_0g\rangle \\ |\phi_0e\rangle \\ |\phi_1g\rangle \\ |\phi_1e\rangle \\
			|\phi_2g\rangle \\ |\phi_2e\rangle \\ |\phi_3g\rangle \\ |\phi_3e\rangle
		\end{matrix} .
	\end{array}
\end{equation}

\begin{figure}
	\centering
	\includegraphics[width=1\textwidth]{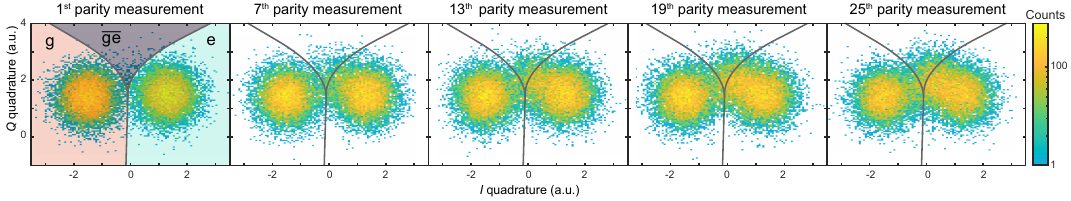}
	\caption{Histograms of single-shot readout signals in I-Q plane of repeated parity measurements. The I-Q plane is divided into three regions. The qubit measured in $\ket{g}$ and $\ket{e}$ is determined by the regions where the readout signal falls. The gray region labeled $\bar{ge}$ corresponds to the qubit being excited to higher energy levels.}
	\label{fig:readout}
\end{figure}

The values of $F_{e\mathcal{G}}$ and $F_{g\mathcal{E}}$ are obtained from the statistical probability of multiple single-shot qubit readout measurements. 
Figure~\ref{fig:readout} shows the I-Q distribution of multiple readout signals for repeated parity measurements. The I-Q plane is divided into three main regions labeled as $g$, $e$, and $\bar{ge}$. The qubit state is determined by the region where the readout signal falls. The qubit state may be disturbed and projected onto energy levels higher than $\ket{g}$ and $\ket{e}$, as the number of repeated parity measurements increases, as shown in Fig.~\ref{fig:readout}. Since it is hard to predict the evolution of the system once the qubit is populated onto a higher excited state, the observed sequences are post-selected, and only those sequences containing no $\bar{ge}$ states are used for HMM analysis.

Given the transition and emission matrices, as well as the observed data, the probability of the initial cavity state being $P(\phi_j)$ can be reconstructed using the forward-backward algorithm \cite{zheng2016accelerating, hann2018robust}, summing over all possible transition paths that start from the cavity state $\ket{\phi_j}$ and yield the observed sequence, with the weight determined by the relevant transition and emission matrix elements:
\begin{align}
	P(\phi_j)=\sum_{S_0\in [|\phi_jg\rangle, |\phi_je\rangle]}\sum_{S_1}\cdots \sum_{S_N}& E_{S_0,R_0}T_{S_0,S_1}E_{S_1,R_1}\cdots T_{S_{N-1},S_N}E_{S_N,R_N}
	\label{eq:bwa}
\end{align}

We then define the likelihood ratio:
\begin{equation}
	\lambda=\frac{P(\phi_1)}{P(\phi_0)+P(\phi_2)+P(\phi_3)}
\end{equation}
to assess the reconstructed cavity state probabilities. A likelihood threshold \( \lambda_{\mathrm{thresh}} \) is then chosen to determine the cavity state in \( \ket{\phi_1} \) if \( \lambda > \lambda_{\mathrm{thresh}} \). A higher threshold exponentially suppresses false-positive errors at a linear cost in the number of measurements.

\subsection{Detector characterization}
Before being used in DP search experiments, the photon detector must be characterized to determine its detection efficiency ($\eta$) and false positive probability ($\delta$). They determine the detector's response to inject a specific number of photons, as described in Eq.~(3) in the main text.

Calibration involves preparing the cavity in the compass state $\ket{\phi_0(\alpha)}$ and applying a coherent drive pulse with varying amplitudes to mimic the displacement operation $D(\beta)$ induced by the DM wave. This operation injects photons into the cavity with an average photon number of $|\beta|^2$. The resultant photon number is measured using repeated QND measurements and HMM analysis, as described in the main text. The measured probability of positive events would increase linearly with $|\beta|^2$. Fitting the data to the model described in Eq.~(3) in the main text yields $\eta$ (slope) and $\delta$ ($y$-intercept). This process is repeated for compass states $\ket{\phi_0(\alpha)}$ with various $\alpha$. 

\begin{figure}[htbp]
	\centering
	\includegraphics[width=1\textwidth]{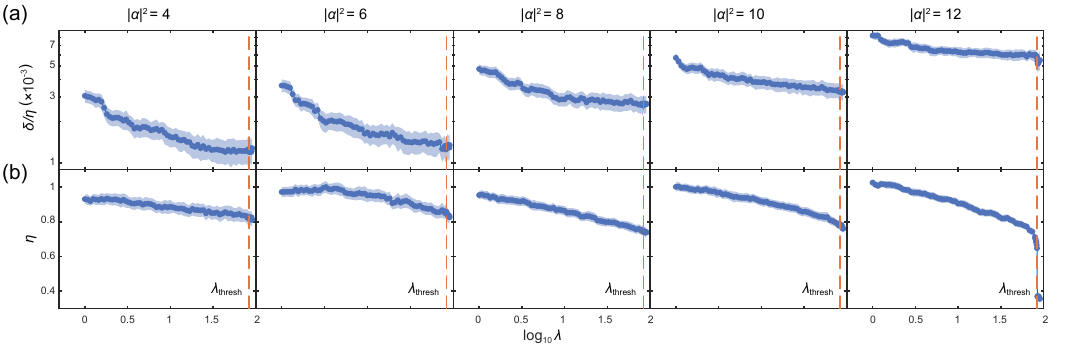}
	\caption{The detector calibration. The normalized false positive probability $\delta/\eta$ (a) and detector efficiency $\eta$ (b) plotted as a function of the likelihood ratio $\lambda$ for compass states with various $\alpha$. The vertical red dashed line represents the chosen likelihood threshold $\lambda_{\mathrm{thresh}}=84$ in the DPDM search experiment.}
	\label{fig:eta_fp}
\end{figure}

The likelihood threshold $\lambda_{\mathrm{thresh}}$ used to distinguish positive and negative events can affect $\eta$ and $\delta$. Both decrease with increasing $\lambda_{\mathrm{thresh}}$, indicating that higher thresholds reduce the false positive rate but also lower the detection efficiency. Figure~\ref{fig:eta_fp} plots the normalized false positive probability $\delta/\eta$ and the detection efficiency $\eta$ as a function of $\lambda_{\mathrm{thresh}}$. The saturation of $\delta/\eta$ at high thresholds suggests that the remaining false positive probability is more likely attributed to some real background populations in the cavity, rather than the overestimation due to qubit- or cavity-based errors. A threshold of $\lambda_{\mathrm{thresh}}=84$ ensures that only events with $\bar{P}_{\phi_1}=1/(1+\lambda)\le 0.0118$ will be counted as positive.

In our experiment, we achieve an 8.1-fold enhancement in the total detection efficiency $\eta_\alpha |\alpha|^2$ for the compass state $|\phi_0(\alpha=\sqrt{12})\rangle$. The decrease of $\eta_\alpha |\alpha|^2$ with increasing $|\alpha|$ is primarily due to additional errors of state preparation and parity measurement arising from increased sensitivity to decoherence, leading to more false-positive events from higher $|\phi_2\rangle$ populations. Because the coherence time of the compass state scales as $T_1^c/|\alpha|^2$, larger $|\alpha|$ leads to more occurrences of errors of state preparation and parity measurement. Additionally, higher-order dispersive and residual couplings contribute to errors in state initialization and measurement. The state preparation errors will increase the $|\phi_2\rangle$ population in the initial compass state, leading to false-positive DP signals. To increase $|\alpha|$ while maintaining high detection efficiency, future technical improvements would focus on: (i) using cavities with higher quality factors and lower intrinsic nonlinearities, (ii) optimizing cavity-qubit coupling to suppress higher-order dispersive effects, and (iii) developing more robust parity-measurement protocols resilient to photon-number-dependent errors.

\begin{table}
	\centering
	\caption{Number of independent trials $N_\mathrm{trials}$ for DP search using both the cat states $\ket{\phi_0(\alpha)}$ with various $\alpha$ and the vacuum state.}
	\begin{tabular}{p{0.8in}|p{0.8in}p{0.8in}p{0.8in}p{0.8in}p{0.8in}p{0.8in}}
		\hline
		\hline
		$|\phi_0(\alpha)\rangle$ & $\alpha=\sqrt{4}$ & $\alpha=\sqrt{6}$ & $\alpha=\sqrt{8}$ & $\alpha=\sqrt{10}$ & $\alpha=\sqrt{12}$ & vacuum\\\hline
		&&&&&&\\[-0.9em]
		$N_\mathrm{trials}$  & 40719 & 35129 & 40621 & 47017 & 28885 & 81040\\
		\hline
		\hline
	\end{tabular}
	\label{tab:Ntrials}
\end{table}

\subsection{Converting background counts to dark photon exclusion}
After the detector calibration, DPDM search can be carried out by measuring the background photon counts in the cavity. Each measurement sequence with a particular value of $\alpha$ for the initial $\ket{\phi_0(\alpha)}$ state is repeated $N_\mathrm{trials}$ times. Positive events $N_\mathrm{meas}$ are determined using the HMM method with an optimized likelihood threshold. The values of $N_\mathrm{trials}$ for both the compass state $\ket{\phi_0(\alpha)}$ with various $\alpha$ and the vacuum state are listed in Table~\ref{tab:Ntrials}. A global fit to the probability of positive events $n_\mathrm{meas}=N_\mathrm{meas}/N_\mathrm{trials}$ as a function of both cat state amplitude $|\alpha|^2$ and the integration time $\tau$ is used to extract the fitting parameters in Eq.~(4) in the main text. The mean and error of these fitting parameters are listed in Table~\ref{tab:fit-para}. Here, both the coefficients $b_{|\alpha|^2}$ and $c_{|\alpha|^2}$ are $\alpha$ dependent. The $b_{|\alpha|^2}$ term arises from incoherent noise and scales linearly with time. This noise may come from the decay through intermediate states $\ket{\phi_i}$ or the residual component $\ket{\phi_2}$ in the initial compass state, which actually increases with $|\alpha|$ and would decay to $\ket{\phi_1}$ during the DM integration process. Since the decay rate increases with $|\alpha|$, we use $b_{|\alpha|^2}$ instead of a constant $b$ in the fitting equation. The false positive probability $c_{|\alpha|^2}$ term comes from the residual component $\ket{\phi_1}$ in the initial compass state, which increases with $|\alpha|$.

\begin{table}[b]
	\centering
	\caption{Fitting parameters and their uncertainties of the DP detection.}
	\centering
	\begin{tabular}{p{1in}|p{1in}p{1in}}
		\hline
		\hline
		Fitted Parameter & $\Theta$ & $\sigma_\Theta$ \\
		\hline
		$a_0(\mathrm{s}^{-2})$ & $3.441\times 10^{4}$ & $1.660\times 10^{4}$\\
		$b_4(\mathrm{s}^{-1})$ & $-2.759$ & $2.036$\\ 
		$b_6(\mathrm{s}^{-1})$ & $0.889$ & $2.103$\\
		$b_8(\mathrm{s}^{-1})$ & $3.542$ & $2.009$\\
		$b_{10}(\mathrm{s}^{-1})$ & $6.913$ & $2.127$\\
		$b_{12}(\mathrm{s}^{-1})$ & $13.215$ & $2.494$\\
		$c_4$ & $1.130\times 10^{-3}$ & $5.438\times 10^{-5}$ \\
		$c_6$ &$1.158\times 10^{-3}$ & $6.032\times 10^{-5}$ \\
		$c_8$ & $1.645\times 10^{-3}$ & $6.654\times 10^{-5}$ \\
		$c_{10}$ & $2.244\times 10^{-3}$ & $7.182\times 10^{-5}$ \\
		$c_{12}$ & $3.142\times 10^{-3}$ & $10.803\times 10^{-5}$ \\
		\hline
		\hline
	\end{tabular}
	\label{tab:fit-para}
\end{table}

After determining the prefactor \( a_0 \) by fitting Eq.~(4) in the main text, the kinetic mixing angle can be calculated using  
\begin{equation}
	\epsilon_0 = \sqrt{\frac{a_0}{\rho_{\mathrm{DM}} m_{\mathrm{DM}} V_{\mathrm{eff}}}} \,
\end{equation}
with the standard deviation determined via error propagation:  
\begin{equation}\label{eq:sig}
	\frac{\sigma_{\epsilon_0}}{\epsilon_0}=\frac{1}{2}
	\sqrt{\left(\frac{\sigma_{a_0}}{a_0}\right)^2+
		\left(\frac{\sigma_{\omega_c}}{\omega_c}\right)^2+
		\left(\frac{\sigma_{V_{\mathrm{eff}}}}{V_{\mathrm{eff}}}\right)^2}
	\approx \frac{\sigma_{a_0}}{2a_0}.
\end{equation}

Experimental parameters for this analysis are provided in Table~\ref{tab:exp-para}. The exclusion bound at 90\% confidence level (C.L.) is given by $\epsilon_{90\%} = \epsilon_0 + 1.28 \sigma_{\epsilon_0}$.
Similarly, exclusion limits at different frequencies are determined using Eqs.~\eqref{eq:agt}, \eqref{eq:gt}, and \eqref{eq:sig}.

\begin{table}
	\centering
	\caption{Experimental parameters and their uncertainties.}
	\centering
	\begin{tabular}{p{1in}|p{1in}p{1in}}
		\hline
		\hline
		Expt. Parameter & $\Theta$ & $\sigma_\Theta$ \\
		\hline
		%		$a_m$ & $1.887\times 10^{5}s^{-2}$ & $0.757\times 10^{5}s^{-2}$\\
		$\omega_c$ & 6.442 GHz &  9.4Hz\\ 
		$V_{\rm eff}$ & 4.45 cm$^3$ & 0.04 cm$^3$\\
		\hline
		\hline
	\end{tabular}
	\label{tab:exp-para}
\end{table}

\subsection{Dark photon search with frequency tuning}
\begin{figure}[b]
	\centering
	\includegraphics{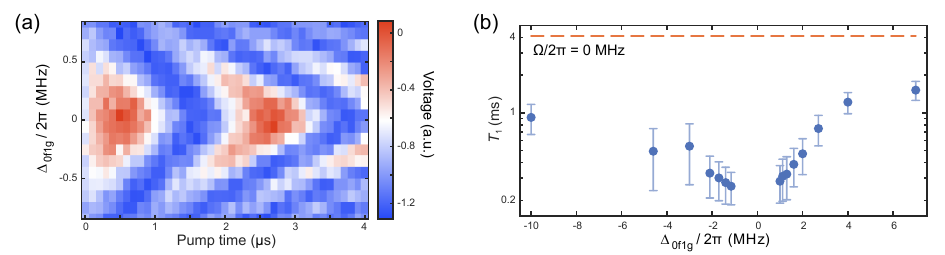}
	\caption{Cavity frequency tuning drive calibration. (a) Rabi Chevron pattern for measuring  $|1, g\rangle$ when preparing $|0,f\rangle$ and switching on the sideband pump with a detuning $\Delta$ for a varying amount of time. The oscillation on resonance gives a Rabi rate of $\Omega/2\pi=0.51$ MHz. 
		(b) Measured cavity coherence time $T_1$ when switching on the sideband drive with different detunings to tune the cavity frequency. The dashed red line indicates the coherence time without the sideband pump.}
	\label{fig:fre_tune}
\end{figure}

The cavity frequency tunability in this work is achieved using a detuned parametric sideband drive $H_d = \frac{\Omega}{2} a^\dagger |g\rangle \langle f| e^{i\Delta t} + \mathrm{h.c.}$, which involves the energy transition between $|n,g\rangle$ and $|n-1,f\rangle$ states of the joint cavity-qubit system. When switching on the parametric drive with $\Delta\approx0$, we observe Rabi oscillations corresponding to the sideband transition $|n-1,f\rangle \leftrightarrow |n,g\rangle$ with a Rabi rate of $\Omega/2\pi=0.51$~MHz for $n=1$, as shown in Fig.~\ref{fig:fre_tune}(a). By engineering a large frequency detuning $\Delta\gg\Omega$, the time-dependent drive Hamiltonian can be approximated to $H_\mathrm{eff} = \frac{\Omega^2}{4\Delta}\left[a^\dagger |g\rangle \langle f|, a|f\rangle\langle g|\right] =  \frac{\Omega^2}{4\Delta} \left(a^\dagger a |g\rangle\langle g| - a^\dagger a |f\rangle\langle f| -|f\rangle\langle f| \right)$. Since the qubit remains in the ground state in our experiment, Hamiltonian terms involving $|f\rangle$ can be ignored, leading to a drive-induced cavity frequency shift $\Omega^2/(4\Delta)$. In our experiment, the cavity frequency is tuned by varying the sideband drive detuning at a fixed Rabi strength, and the measured cavity frequency shift as a function of the sideband drive detuning is shown in Fig.~4(b) of the main text. This result agrees well with the theoretical prediction, demonstrating tunability of the cavity frequency over about 100~kHz. We also investigate the cavity coherence time as the cavity frequency is tuned. The measured cavity coherence $T_1$ time is plotted as a function of the sideband drive detuning in Fig.~\ref{fig:fre_tune}(b), as compared to the $T_1$ without applying the sideband pump. The measured results indicate a reduction in the cavity coherence time when the frequency-tuning pump is switched on with a small detuning due to the breakdown of the approximation $\Delta\gg\Omega$.  

To perform DP search experiments while tuning the cavity frequency, we first characterize the DP detector with an initial vacuum state using a similar HMM analysis~\cite{Dixit2021} at each frequency. In this scenario, the joint cavity-qubit hidden states are given by $S\in [|0g\rangle, |0e\rangle, |1g\rangle, |1e\rangle]$, and the transition and emission matrices now read as
\begin{equation}
	T=\begin{pmatrix}
		P_{00}P_{gg} & P_{00}P_{ge} & P_{01}P_{ge} & P_{01}P_{gg} \\
		P_{00}P_{eg} & P_{00}P_{ee} & P_{01}P_{ee} & P_{01}P_{eg} \\
		P_{10}P_{gg} & P_{10}P_{ge} & P_{11}P_{ge} & P_{11}P_{gg} \\
		P_{10}P_{eg} & P_{10}P_{ee} & P_{11}P_{ee} & P_{11}P_{eg} 
	\end{pmatrix}, 	
\end{equation} 
and
\begin{equation}
	E=\frac{1}{2}\begin{pmatrix}
		F_{g\mathcal{G}} & F_{g\mathcal{E}} \\
		F_{e\mathcal{G}} & F_{e\mathcal{E}} \\
		F_{g\mathcal{G}} & F_{g\mathcal{E}} \\
		F_{e\mathcal{G}} & F_{e\mathcal{E}} 
	\end{pmatrix}.
\end{equation} 

\begin{figure}
	\centering
	\includegraphics{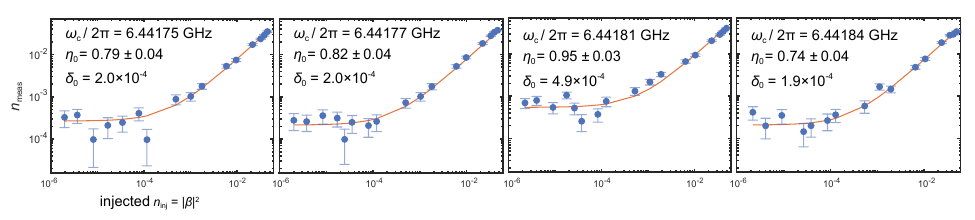}
	\caption{DM detector calibration at different cavity frequencies with a chosen threshold $\lambda_{\mathrm{thresh}}=10^5$.}
	\label{fig:dec_cal}
\end{figure}

By using a similar forward-backward algorithm in Eq.~\eqref{eq:bwa}, we can reconstruct the cavity state probabilities $P(n_0=0)$ and $P(n_0=1)$, and define the likelihood ratio $\lambda=P(n_0=1)/P(n_0=0)$ to determine the cavity photon number. Similarly, by defining a likelihood threshold, we can extract the probabilities of positive events for measuring the single-photon state as a function of the injected photon number $|\beta|^2$. The experimental results at different cavity frequencies are presented in Fig.~\ref{fig:dec_cal}, which are fitted to a linear function of $n_{\text{meas}} = \eta n_{\text{inj}} + \delta$ to extract the detection efficiency $\eta$ and the false positive probability $\delta$ at each cavity frequency, as shown in Fig.~4(c) in the main text.  

\begin{figure}[b]
	\centering
	\includegraphics{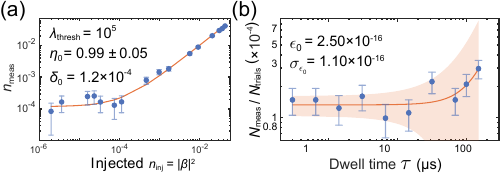}
	\caption{DP search using vacuum states. (a) The detector calibration results. (b) Measured probability of positive events as a function of signal integration time in the DP search experiment.}
	\label{fig:vacuum}
\end{figure}

After the detector calibration, we conduct the DP search experiment with a signal integration time $\tau$ at each tuned cavity frequency. Here, $\tau$ is engineered to match the cavity's coherence time at each frequency [Fig.~\ref{fig:fre_tune}(b)] to maximize the accumulated DP-induced signal. Note that the cavity's coherence time is much longer than the DM coherence time, while the DM is an incoherent source and cannot be distinguished from the background noise at a single frequency. We then determine the background occupation at each frequency. As a comparison, we first perform a DP search experiment without applying the frequency-tuning pump, and present the results in Fig.~\ref{fig:vacuum}.

In the DP search experiment as varying the cavity frequency, the tuning step is set to $\Delta\omega\approx 6$ kHz to match the DM linewidth, ensuring that the DM signal appears at most in one of 16 frequency bins. Assuming that the background noise remains unchanged within the 16 bins, we perform a constant fit. This introduces an attenuation factor $\eta_{\mathrm{fit}}$ to the potential signal. Any signal entering the fit function will be reduced by a factor of $1/16$, resulting in $\eta_{\mathrm{fit}}=93.75\%$.

\begin{figure}[t]
	\centering
	\includegraphics{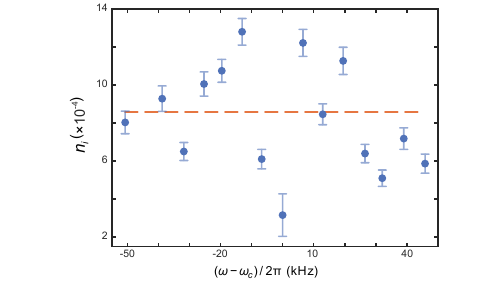}
	\caption{Measured background occupations at each tuned cavity frequency, with the error bars showing only statistical errors.}
	\label{fig:background}
\end{figure}

The background occupation at each frequency \(\omega_i\) is defined as  
\begin{equation}
	n_i = \frac{N_{\text{mea},i}}{\eta_i N_{\text{trial},i}},
\end{equation}  
where \(N_{\text{mea},i}\) is the number of excitations measured, \(N_{\text{trial},i}\) is the number of trials, and \(\eta_i\) is the detection efficiency. The background is estimated using a weighted average and variance:  
\begin{equation}
	\bar{n} = \frac{\sum_i N_{\text{trial},i} n_i}{\sum_i N_{\text{trial},i}}, \quad \sigma_n^2 = \frac{\sum_i N_{\text{trial},i} (n_i - \bar{n})^2}{\sum_i N_{\text{trial},i}}.
\end{equation}
and plotted in Fig.~\ref{fig:background}. The signal rate is defined as  
\begin{equation}
	p_i = \frac{n_i - \bar{n}}{n_{\text{ref},i}} \quad\mathrm{with}\quad n_{\text{ref},i} = \eta_{\text{fit}} n_{\text{sig}}(m_{\rm{DM}}, f_i, \epsilon = 1),
\end{equation}  
where the reference signal rate \(n_{\text{ref},i}\) accounts for the attenuation factor \(\eta_{\text{fit}}\). The signal power \(n_{\text{sig}}\) is given by  
\begin{equation}
	n_{\text{sig}} = \epsilon^2 \rho_{\mathrm{DM}} V_{\text{eff}} m_{\mathrm{DM}} T^1_{c,i} \mathcal{F}(m_{\mathrm{DM}}, \omega_i),
\end{equation}  
where \(\mathcal{F}(m_{\mathrm{DM}}, \omega_i)\) is the normalized frequency spectrum of the DPDM, and we neglect the decoherence uncertainty by choosing the integration time to be $T^c_{1,i}$ in the $i$-th frequency bin.

Using error propagation, the variance of the signal rate is calculated as  
\begin{equation}
	\sigma_{p_i}^2 = \left(\frac{\sigma_{n_i}}{n_{\text{ref},i}}\right)^2 + \left(\frac{n_i - \bar{n}}{\sigma_{n_i}}\right)^2 \left[\left(\frac{\sigma_{T^c_{1,i}}}{T^c_{1,i}}\right)^2 + \left(\frac{\sigma_{V_{\text{eff}}}}{V_{\text{eff}}}\right)^2\right]\approx \left(\frac{\sigma_{n_i}}{n_{\text{ref},i}}\right)^2.
\end{equation}
which is dominated by the statistical uncertainty from the background estimation.

Since the reference signal \(n_{\text{ref},i}\) is defined for \(\epsilon = 1\), the probability function can be expressed in terms of \(\epsilon\):  
\begin{equation}
	P(p_i | \epsilon) = \frac{1}{\sqrt{2\pi} \sigma_{p_i}} \exp\left(-\frac{(p_i - \epsilon^2)^2}{2 \sigma_{p_i}^2}\right) / \text{Const},
\end{equation}  
where \(\text{Const}\) is a normalization factor ensuring \(\int_0^\infty P(p_i | \epsilon) d\epsilon^2 = 1\). %The product is over nearby resonant bins, as scans are separated by \(\approx 6~\text{kHz}\), corresponding to \(m_{\mathrm{DM}} / (2\pi Q_{\mathrm{DM}})\).

The \(90\%\) upper limit on the kinetic mixing coefficient \(\epsilon\) is obtained by inversely solving  
\begin{equation}
	\int_0^{\epsilon_{90\%}^2} P(p_i | \epsilon) d\epsilon^2 = 90\%.
\end{equation}

\end{document}